\pdfoutput=1
\documentclass[twoside]{article}

\usepackage{lipsum} 

\usepackage[sc]{mathpazo} 
\usepackage[T1]{fontenc} 
\usepackage{microtype} 

\usepackage[hmarginratio=1:1,top=32mm,columnsep=20pt]{geometry} 
\usepackage{multicol} 
\usepackage[small,up,textfont=up]{caption} 
\usepackage{booktabs} 
\usepackage{float} 
\usepackage{hyperref} 

\usepackage{lettrine} 
\usepackage{paralist} 

\usepackage{abstract} 

\usepackage{titlesec} 
\renewcommand\thesection{\Roman{section}} 
\renewcommand\thesubsection{\Roman{subsection}} 
\titleformat{\section}[block]{\large\scshape\centering}{\thesection.}{1em}{} 
\titleformat{\subsection}[block]{\large}{\thesubsection.}{1em}{} 

\usepackage{fancyhdr} 
\usepackage{graphicx}
\usepackage{epstopdf}
\usepackage{epsfig}
\usepackage{natbib}

\newcommand{\sgn}{\mathop{\mathrm{sgn}}}

\title{\vspace{-15mm}\fontsize{24pt}{10pt}\selectfont\textbf{Strong random correlations in networks of heterogeneous agents}} 

\author{
\large
\textsc{Imre Kondor$^{1,2}$}\thanks{Corresponding author. E-mail: kondor.imre@gmail.com}
\textsc{,Istv\'an Csabai$^{1}$,}
\textsc{G\'abor Papp$^{1}$,}\\
\textsc{Enys Mones$^{1}$,}
\textsc{G\'abor Czimbalmos$^{1}$,}
\textsc{M\'at\'e Csaba S\'andor$^{1}$}
\\[2mm]
\normalsize $^1$E\"otv\"os Lor\'and University, Budapest, Hungary\\ 
\normalsize $^2$Parmenides Foundation, Pullach b. M\"unchen, Germany\\ 
\vspace{-5mm}
}
\date{}

\begin{document}

\maketitle 

\thispagestyle{fancy} 


\begin{abstract}

Correlations and other collective phenomena are considered in a schematic model of pairwise interacting, competing and collaborating agents facing a binary choice and situated at the nodes of the complete graph and a two-dimensional regular lattice, respectively. The agents may be subjected to an idiosyncratic or common external influence and also some random noise. The system's stochastic dynamics is studied by numerical simulations. It displays the characteristics of punctuated, multiple equilibria, sensitivity to small details, and path dependence. The dynamics is so slow that one can meaningfully speak of quasi-equilibrium states. Performing measurements of correlations between the agents' choices we find that they are random both as to their sign and absolute value, but their distribution is very broad when the interaction dominates both the noise and the external field. This means that random but strong correlations appear with large probability. In the two dimensional model this also implies that the correlations on average fall off with distance very slowly: the system is essentially non-local, small changes at one end may have a strong impact at the other. The strong, random correlations tend to organize a large fraction of the agents into strongly correlated clusters that act together. If we think of this model as a metaphor of social or economic agents or bank networks, the systemic risk implications of this tendency are clear: any impact on even a single strongly correlated agent will not be confined to a small set but will spread, in an unforeseeable manner, to the whole system via the strong random correlations.

\end{abstract}


\begin{multicols}{2} 

\section{Introduction}
\label{sec_int}

The view of the economy as a complex adaptive system pioneered by \citet{1} and developed in \citet{2}, \citet{Art97} and a large number of subsequent papers is gaining increasing significance in economics. An important approach to such a complex adaptive system is offered by agent based modelling (ABM), \citet{3}. In its ambition to overcome the unrealistic simplifications of general equilibrium theories, it tries to model the behaviour of the individual agents by taking into account their heterogeneity, and to generate the phenomena observed on the macro level as the manifestations of the collective behaviour of these heterogeneous interacting agents. As a consequence of the wide variety of the characteristics of the agents and their interactions, ABM's are necessarily very high dimensional: they depend on a large number of variables. The calibration and validation of these models is therefore a very difficult endeavour, which is the source of frequent, and justified criticism, see e.g. \citet{4}. A closely related issue is the high degree of instability of these models: very small modifications of the details may lead to unforeseeable changes in the overall behaviour. The high dimensionality of agent based models may be seen either as a damning fault or a virtue, depending on whether one believes that knowledge about society or the economy should be possible to be compressed into a few postulates, or is willing to accept the irreducible complexity of economic and social activity.

Agents are often represented as the nodes of a complex network or graph, and their interactions as the links of this network. In a typical setup, the agents have idiosyncratic goals and specific strategies to reach them, and they are also invested with the capability of learning and adaptation. This is a great virtue of the ABM approach, but further aggravates the calibration and the instability problem. In this paper we are going to consider a hugely simplified version of such a model: binary agents having a simple yes-no alternative in front of them, and linked to each other by fixed interactions, symmetric in the interacting pair. So our agents do not learn, but try (according to a probabilistic rule) to react to the influence of their partners, and possibly also to that of an external source. This model retains an important aspect of ABM's in that the interactions between the agents are supposed to be heterogeneous. However, for the sake of numerical simplicity, we assume that even this heterogeneity consists merely in some of the interactions being positive (friends or collaborators), some negative (enemies or competitors), but the absolute values are the same. (A richer realization of the interactions, such as drawing them from a continuous probability distribution, could be easily accommodated.)

This model is the same as the standard model used to describe a class of disordered magnets called spin glasses, \citet{5}. The way we use the model will however be rather different from the approach of statistical physics; the differences will be explained later.

Social interaction models inspired by statistical physics, sometimes especially spin glasses, have been proposed in the context of various socioeconomic problems for quite some time now, \citet{Dur96}, \citet{Bro01}. \citet{Dur99} writes:

``\dots this approach introduces an explicit sociological perspective on individual behavior. By moving away from market-mediated interrelationships to direct interdependencies in behavior, this new approach to analysis represents an ''interactions-based'' approach to socioeconomic behavior in which individual decisions are explicitly understood as determined by one's social context. A large class of these interactive decisions are binary: Staying in or dropping out of school, the decision to have an out-of-wedlock birth, the use of drugs, and entry into illegal activity all have this characteristic.''

Sequential voting in a two-party election system, e.g. \citet{Bar12}, can also be added to the above examples.

In a different context \citet{Kru94} argues that in order to model  complex  landscapes  in  economic  geography,  a  useful  metaphor  is  provided  by  spin-glass models. Incidentally,  he  notes: ``The  residential  segregation  model  introduced  [...] by  \citet{Sch78} bears a strong resemblance to simple spin-glass models.'' 

Spin glasses appear also in the context of rational decision making, \citet{Gal98}, portfolio selection under nonlinear constraint, \citet{Gab99}, optimization of capital charge under international financial regulation, \citet{Kon00}, coalition formation and fragmentation, \citet{Gal08}, in the dynamics of social networks, \citet{Sha05}, or the description of financial markets, \citet{Ros02}, and \citet{Bur13}.

Our goal in this paper is not to describe another application of spin glass theory to a concrete social or economic problem, but to display the remarkable richness of behaviour this family of models can exhibit, including the sensitivity to fine details and emergence of long range correlations that make the system ``more than the sum of its parts''.

Given the socioeconomic context of this work, we deviate from the standard statistical physics approach in some important respects. We do not seek to reach very large system sizes because the effects we wish to exhibit can be demonstrated already on moderate sized systems. Furthermore, most of the applications we have in mind are concerned with systems that would qualify at most as mesoscopic, rather than macroscopic from the point of view of physics.   Except for the smallest sizes, we do not necessarily attempt to equilibrate our samples either. The reason is that heterogeneous agent models relax extremely slowly after a shock, and the next shock may well arrive before they could settle down into equilibrium. A recent paper by \citet{Gua13} gives an excellent demonstration of this type of behaviour -- in perfect agreement with what we perceive as the true dynamics of markets.

To consider the limit of infinitely large systems is well justified in physics that is concerned with macroscopic systems consisting of, say, $10^{24}$ particles. In the limit of large particle numbers the macroscopic thermodynamic quantities self-average, i.e. become independent of the realization of the interaction matrix. This allows one to average over the random samples (quenched averaging), which simplifies the treatment tremendously. It must be stressed, however, that not all quantities become insensitive to the details of the interactions: the spin-glass overlap and two-spin correlations, for example, show random variations from sample to sample, \citet{5}, \citet{9}. (The importance of the lack of self-averaging has been a\-na\-lysed in the context of the relationship between micro- and macroeconomics by \citet{Aok10}, \citet{Aok12}, and \citet{Gar12}.) One may feel that such sample dependent, randomly varying quantities carry no meaningful information. We wish to show, however, that some typical features do emerge, and they must be explored if we wish to associate these models with some real world systems, such as a group of human actors, or a network of institutions.

One of the typical features we observe is that a large cluster of strongly correlated agents is formed for almost any realization of the interaction matrix. We believe that strong correlations are a fundamental feature of complex systems, but we are not aware that this would have been emphasized in the context of agent based models. From the point of view of physics, there is nothing mysterious about the emergence of strong correlations: they are the direct consequence of the interaction between the agents, and the manifestations at the mesoscopic level of what would become a genuine phase transition with its associated long range order in a macroscopic system. We argue that the high sensitivity to small details of the models studied in this paper is a consequence of the presence of the large correlated clusters. This sensitivity means these systems are effectively \textit{irreducible}: they depend on so many details that a reduced description in terms of a small number of variables is impossible. (\citet{10} once gave a deceptively innocent looking definition of a complex system as one whose behavior crucially depends on small details.) Another aspect of the same property is the \textit{nonlocal} character of these systems. With the large strongly correlated clusters spanning the whole system, a local disturbance at one point may have an effect which extends to the whole system.

A further typical feature is that these systems exhibit a complicated attractor structure: depending on the actual arrangement of the interactions between the agents, the system's evolution may be arrested in many more or less deep ''valleys'' (basins of attraction) in which the system will be trapped for long time periods before it escapes via lower or higher passes, only to be trapped in the next valley. This induces an extremely slow, punctuated equilibrium-like dynamics which is quite reminiscent of the quasi-equilibria separated by regime changes in biosystems or societies or markets. The existence of many attractors makes the system's dynamics dependent on initial conditions: trajectories started at different microscopic configurations may end up in different basins of attraction, and even trajectories which are initially very close may evolve very far from each other. In other words, we have strong path dependence in these models. Moreover, even the landscape itself may be strongly rearranged in response to small changes in the interaction structure, in the control parameters or in the boundary conditions.

This leads us to the problem of equilibration. Heterogeneous, glassy \textit{physical} systems may take an extremely long time to equilibrate. As a matter of fact, laboratory spin glasses never reach true thermodynamic equilibrium, i.e. a state in which \textit{no macroscopic quantity would change any more, and all the previous history would be forgotten}\footnote{Note how different this concept of equilibrium is from that in economics.}.

The numerical simulation of sufficiently large spin glass models in physics is a challenging task that requires tremendous numerical effort and programming ingenuity. The context of the present work is, however, entirely different: the social, economic or finance systems for which our schematic binary agent model can be considered at least as a metaphor are even less likely to come into equilibrium than their spin glass counterparts, therefore we do not need try to reach ideal equilibrium in our simulations. Sometimes we will consider, mainly for the sake of illustration, small enough systems that allow us to scan the whole of phase space (i.e. the complete set of microscopic configurations) and compare the results thus obtained with the results extracted from Monte Carlo simulations. Generally, however, we will adopt the position that our Monte Carlo simulations are a kind of history of the artificial society of interacting agents, and content ourselves with following this history for a sufficiently long time to draw some conclusions. The slow dynamics and the existence of quasi-equilibrium states allow us to measure various quantities by averaging over long, but finite periods of time. 

In the context of statistical physics, going to the limit of very large system sizes and very long time spans is not only natural, but also a powerful means of reduction: many small details lose their significance and become forgotten along the way. Accordingly, the results will be stable and smooth. In renouncing the thermodynamic limit and equilibrium as powerful tools of simplification, we are evidently sacrificing some reproducibility (results will depend on the sample, on initial and boundary conditions, etc.), also some of the beauty of smooth results, and may also run into difficulties of interpretation. As an example of the latter, we mention that whereas in a large, equilibrium system the distinction between a stable structure and fluctuations about it is clear-cut, in finite size, out of equilibrium systems it is blurred. On the other hand, our ''mesoscopic'' approach also has some benefits: we are relieved of the impossible task of simulating huge systems for infinitely long times. 

We know from statistical physics that the nature, in fact even the existence, of collective phenomena and ordering depend on the underlying topology. In this paper we are going to consider two kinds of underlying geometries: the complete graph and $2$-dimensional regular lattices. Agents occupying the nodes of a complete graph directly interact with every one of their peers: this could correspond to a close-knit group. In contrast, agents occupying the lattice sites on, say, a square lattice form something like a sparsely populated flat country: agents have only a small number ($4$) of partners with whom they are directly linked. In a sense, these cases represent two extremes: on the complete graph the distance between any pair of agents is $1$, everybody is nearest neighbour to everybody else. In contrast, on a $2$-dimensional regular lattice of $N$ lattice sites the average distance between two randomly chosen agents is of the order of $N^{\frac{1}{2}}$, i.e. it grows rather fast with system size. More realistic multi agent models would be implemented on some kind of complex random graphs. The structure of these falls somewhere in between the above two cases: complex random graphs tend to display the small world property, \citet{Wat98}, with the average distance between agents growing as the logarithm of the system size.

It is rather natural to expect the direct coupling between each pair of partners on the complete graph to induce strong correlations between them, and we will indeed find that large, strongly correlated clusters readily emerge. It is less obvious that the same phenomenon should arise in low dimensional lattices, but we will see that extended correlated clusters, spanning the whole sample build up in low dimensional lattices too. A similar study of correlations on complex random graphs poses no difficulty in principle, but would introduce, in addition to the random distribution of the interactions, further elements of complication, related to the random network on which the agents would be placed. Therefore we decided to postpone the analysis of correlations on random graphs to a later work, but there is no doubt in our minds that these graphs, falling between the two extremal structures of low dimensional lattices and the complete graph, will also be found to yield extended correlations. The same is true also for extending the model in other directions, such as allowing a richer structure for the interactions between agents and/or for the set of choices available for them.

To summarize, our approach consists in studying the behaviour of not necessarily large (sizes going from $\mathcal{O}(1)$ through $\mathcal{O}(100)$, in a few exceptional cases up to $\mathcal{O}(10^4)$) schematic  (spin glass-like) heterogeneous agent systems on various geometries, for times that are not necessarily long enough to reach equilibrium, and looking for regularities, typical features in the idiosyncratic and casual behaviour of these miniature societies of binary agents. 

The paper is organized as follows. In Section~\ref{sec_model} we specify the model and lay down the rules of its dynamics. In Section~\ref{sec_dynamics} we show the results of some long Monte Carlo simulations that follow the evolution of a relatively large, $N=10^4$, sample and display the type of slow dynamics where the results depend on the length of observation time. In Section~\ref{sec_phasespace} we display a few examples of the cost function landscape and the network of low lying (low cost) states (very different from, but depending on, the network of interactions) on which the dynamics unfolds. Section~\ref{sec_correlations} gives a description of the correlations emerging between the agents on the complete graph in the absence resp. presence of an external field, while Section~\ref{sec_lattice} displays correlation results in two-dimensional regular lattices. Section~\ref{sec_boundary} demonstrates the role of boundary conditions. Section~\ref{sec_summary} is a summary of the results.

\section{The model} 
\label{sec_model}

Let us consider a system of $N$ agents, labelled by $i=1,2,\dots,N$. They are faced with a binary choice, the choice agent $i$ makes will be denoted by $s_i = \pm1$. The microscopic states of the system are given by $N$-vectors with components $\pm 1$. In all, there are $2^N$ such vectors; they form a hypercube in $N$-dimensional space. The complete set of these microscopic states constitutes what in physics is called the ''phase space'' of the system.

The agents interact with each other via a two-body interaction $J_{i,j}$. This interaction can be chosen to be binary again: $J_{i,j} = \pm1$, according to whether agents $i$ and $j$ are friends or enemies, cooperate or compete with each other. The interactions are supposed to be symmetric: $J_{i,j} = J_{j,i}$ . Any specific choice of the interaction or coupling matrix $J_{i,j}$ defines a given realization or \textit{sample} of the model. The binary distribution of interactions could be replaced, within limits, by some other distribution without changing the main message of this paper. A frequently used alternative assumes that the couplings are drawn from a Gaussian distribution. For the sake of numerical convenience, we will keep to the binary interaction matrix.

As mentioned above, the agents can be placed on the complete graph (each agent having a direct link to every other), or on a regular Euclidean lattice (such as the square lattice in two dimensions). Random, complex networks will be considered in a subsequent paper. 

The dynamics of the system is governed by a cost function. In the physics context, this cost function is the energy functional or Hamiltonian. In our present case, its meaning depends on the interpretation one assigns to the agents and their interactions. A few examples will be given at the end of this section. For the time being, it may be useful to think of this cost as a social discomfort arising from the competing interactions between the agents. For a given realization of the interactions and in a given microscopic state the cost is given as follows:

\begin{equation}
E=-\sum\limits_{\left<i,j\right>}J_{i,j}s_is_j-\sum\limits_{i}h_is_i\\ \label{eq_ham_1}	
\end{equation}

The first sum is meant to be over the interacting $i,j$ pairs. The second term in Eq.~(\ref{eq_ham_1}) describes the coupling of the agents to the external ''fields'' $h_i$. These fields may be different for each agent, or the same for all of them:  $h_i \equiv  h$. The fields are supposed to be real (positive or negative) numbers. (Possible interpretations of these parameters will be mentioned at the end of this section.)

We assume that, in order to decrease the social discomfort, the agents tend to make the same choice as their friends and the opposite to their enemies. At the same time, they try to align with the external field (i.e. choose $s_i = \sgn(h_i)$~) acting on them. All this amounts to their trying to decrease the value of the cost function. Flipping the values of the $s_i$  induces a walk in phase space: updating an agent's status moves the state vector into a neighbouring corner of the hypercube. We assume that this walk is performed in consecutive steps, where randomly selected agents serially update their states according to the influence of their peers and of the external field, and keep or reject the new, updated choice according to whether it decreases or increases the cost given in Eq.~(\ref{eq_ham_1}).

What we have described so far is a random search for the minimum of the cost. The task can be complicated by two kinds of conflicts that are built into the model by design. One of these is the potential conflict between the field $h_i$ acting on a given agent $i$ and the influence from the side of her peers: The agent may have a strong conviction that the proper choice is $+1$ (modelled here by her being subjected to a strong positive field $h_i$), but the majority of her friends may have chosen $-1$. Another source of internal conflict is the fact that the interactions between the agents may not be transitive: if agent $i$ is a friend of agent $j$ and $j$ is a friend of $k$ this does not imply that $k$ will be a friend of $i$. If $k$ and $i$ happen to be enemies, the trio $i$, $j$, $k$ will not be ''happy'' for any configuration of their choices, one of the bonds $J_{i,j}$ , $J_{j,k}$ , $J_{k,i}$ will not be satisfied for any values of the $s_i$, $s_j$, $s_k$. Such a trio is called \textit{frustrated}. It can be seen that whenever the product of the couplings around a closed loop is negative, frustration will appear, \citet{15}. 

The internal conflicts built into the model lead to two important consequences: On the one hand, the cost landscape will typically be very rugged, displaying a large number of local minima, with a nearly or completely identical value of the cost at several of these; the optimization task, for a generic choice of the fields and couplings, is strongly non-convex. On the other hand, in a typical instance of the problem even the minimal achievable costs are much higher than they would be without these internal conflicts.

The rugged cost landscape means that a greedy algorithm which accepts an update only when the cost decreases will have a high chance of getting trapped in one of the several local minima, possibly very far from the true minimum. The number of local minima grows very fast with increasing system size.

In order to endow our agents with a certain degree of flexibility, we assume that, in addition to their mutual influence on each other and the external field acting on them, the agents are also exposed to some noise that we characterize by what physicists would refer to as a ''temperature'' and which we denote by $T$. This is very common in economics and in particular in game theory, where individuals only adopt the best response or the optimal choice with a certain probability. The impact of this noise is that sometimes it makes the agents fail to follow the influence of their partners and the field. If the update of the state of a given agent results in the decrease of the value of the cost function or leaves it unchanged, we accept that move. If the update increases the cost, we accept it with probability

\begin{equation}
p=e^{-\frac{\Delta E}{T}} \\	
\label{eq_mmc}
\end{equation}	
and reject it with probability $1-p$. At high noise levels (high $T$)  good ($\Delta E < 0$) or bad ($\Delta E > 0$) moves are accepted with nearly equal probability, therefore any structure that would be favoured by the interactions or the fields is washed away, while at low noise levels the interaction dominates the noise, and bad moves will be rejected with a high probability.

The dynamics described above is a standard serial Metropolis Monte Carlo algorithm. The acceptance rule in Eq.~(\ref{eq_mmc}) has been imported from statistical physics. There, its choice is motivated by the fact that in the long time limit it leads to the Boltzmann distribution, the equilibrium distribution in a physical system. As explained in the Introduction, we do not wish to mimic statistical physics, in particular, we do not necessarily wish to reach equilibrium, so for us Eq.~(\ref{eq_mmc}) is just a convenient choice to simulate the history of this set of agents.

There exist alternatives to the dynamics described above, some of them much more sophisticated and efficient (in the sense that they bring the system into equilibrium faster) than the one above. \footnote{ An obvious example is the following: If the agents occupy the nodes of a bipartite graph, they can be divided into two subsets such that every agent belonging to subset A is directly linked only to agents belonging to subset B and vice versa. Then each subset can be updated simultaneously, still applying the update rule Eq.~(\ref{eq_mmc}), which evidently results in a tremendous gain in simulation time. We have actually used this method in the case of the square lattice, and checked that it leads to the same result as the sequential update, provided the latter is run for a \textit{large number of sweeps}. An arbitrary redefinition of the rules of the game may, of course, lead to arbitrarily different results, including the flip-flop type of behaviour if one decided to update \textit{every} agent simultaneously. } We will keep to the dynamics generated by Eq.~(\ref{eq_mmc}), partly because of its simplicity, partly because it can be seen as the agents reacting individually to their actual environment, yet achieving some common ''good'', the decrease of the cost.

A careful and detailed discussion of why and how the machinery of statistical physics can be applied to socioeconomic problems can be found in \citet{Dur96, Dur99}, \citet{Bro01}, \citet{Aok98}, \citet{Con07}, and \citet{Agl10}, among others.

The authors just quoted and some of those cited earlier demonstrated that statistical physics methods, concepts and models like the one treated here can prove to be useful in analysing some social and economic phenomena. As we have already remarked, we do not wish to consider concrete applications here. Nevertheless, in order to help intuition and anchor ideas, we suggest to keep some real life example in mind when assessing the results in the following sections. Voting in a two-party political system constitutes such an example. Here agents do have a binary choice indeed, and their decisions are obviously influenced by their relationships to other agents. Furthermore, their decisions may also depend on some idiosyncratic factor like family tradition or an inner conviction that may act as a local field $h_i$ and may compete with the influence of their peers. Alternatively, the whole society of agents may be subject to indoctrination or intimidation that would act on them as a homogeneous field $h$. Finally, the role of noise would be played here by disillusionment, drunkenness, or any other factor that might cause the agents to cast their vote at random, irrespective of the opinion of their peers or any field-like factor. Obviously, other binary choices like buying a fashionable item or not, dropping smoking or not, joining the Libor collusion or not and countless others where the influence of the social environment and external factors (marketing propaganda, health campaigns, etc.) has an impact on the agent's decision can be approximately described by similar models.

There are, of course, some difficulties with this interpretation: voting usually does not take place in long consecutive series (although serial voting is sometimes considered in the literature, see \citet{Bar12}), but we can think of repeated opinion polls or the long series of primaries in the US presidential campaing which reveal for the agents the probable position of others and allow them to update their own position several times. This example  also shows very clearly why running the simulation of this model too long may sometimes be completely superfluous.

To conclude this section, we wish to provide some preliminary considerations about the behaviour of the model in Eq.~(\ref{eq_ham_1}) for different choices of the interaction matrix $J_{i,j}$. For simplicity, let us consider now the case when no external field is present. Then it is clear from Eq.~(\ref{eq_ham_1}) that a simultaneous flip of the signs of all the $s_i$'s does not change the cost function: without the fields the problem has an overall symmetry. This symmetry is exact, and the dynamics generated by Eq.~(\ref{eq_mmc}) also respects it. This might lead us to believe that mirror image microscopic configurations will appear with equal probability, wherever we happen to start the simulation. This is indeed so, provided the noise level is high enough. For $T\gg|J|=1$ the acceptance rule hardly differentiates between steps that increase or decrease the cost, the phase space trajectory can freely roam the complete set of microscopic states. This means that the system is ergodic, averaging over a long enough trajectory in time or over the ensemble of all microscopic states gives identical results.

Assume now that we gradually reduce the noise level, $T$ becoming of order of $1$. Then the interaction starts to make itself felt, steps reducing the cost will be definitely more probable than those that increase it, the system starts to locate those regions in phase space that are at or around the lowest cost states. Because of the symmetry just mentioned, these minimal cost states must come in pairs, differing from each other in that all the $s_i$'s change sign. In the special case of all the couplings being positive, these minimal cost states can be readily identified: either all agents vote $+1$, or all of them vote $-1$. Any other state has a higher cost, and in order to go from one of the minimal cost states to the other we have to flip all the $s_i$'s. The path leading from the ''all $+1$'' state to the ''all $-1$'' state will necessarily go through regions where roughly half the agents choose $+1$ and half of them choose $-1$, and at low $T$ these are among the highest cost states. 

Assume now that $T$ is small, much smaller than $1$. Then the probability of a bad move that increases the cost is very small. If we start the simulation in the basin of attraction of one of the minimal cost states, it will be extremely improbable that the trajectory can climb over the barrier separating the two basins of attraction, especially when the system is large that makes the barrier very high. Therefore the system will be trapped in that half of phase space where it started its random walk. As a result, the time average will no longer be equal to the ensemble average, that is ergodicity will be broken.

The phenomenon we have just described is a simplistic picture of how a ferromagnetic phase transition takes place in a system where the atomic magnetic moments tend to align with each other. Translated into the language of a serial voting process, we can say the following. If everybody casts their vote at random (high noise regime), then the outcome will be determined by the law of large numbers, and will be nearly a draw with high probability. On the other hand, if agents strongly cooperate (interaction dominated regime), everybody trying to vote as their peers, but nobody having any specific preference as to which way to vote (because the external fields have been switched off, so the system is symmetric to a simultaneous flip of all the votes), then the outcome will be a unanimous $+1$ or a unanimous $-1$, determined by any small asymmetry in the initial condition, or even a single voter having a preference. Note that in this case, the magnitude of the asymmetry does not really matter, the system ''wants'' to order, and is willing to follow a leader even if her weight is very small at the beginning.

The case of all the couplings being the same $+1$ corresponds to a hopelessly conformist society where everybody aligns with everybody else, and which is therefore an easy prey to even an insignificant force, immensely amplifying its effect by aligning with it \textit{en mass}. Let us now examine a different structure. If we choose a sample from the set of interaction matrices at random, we will likely select a sample having about the same number of positive and negative couplings. In the high noise regime this sample will not behave any differently from the previous one, the noise supresses the interactions. At low noise levels the picture will be very different, however. Interactions will dominate, but these interactions are in internal conflict now, the system is highly frustrated. The cost landscape will now display not two, but a large number of valleys, each of them having their mirror image pair, but otherwise more or less unrelated. There will be barriers between them, and at sufficiently low noise levels the system will be trapped in one of these valleys. Translated to the voting model again: the outcome of voting in a 50-50 \% polarized, partisan society can be close to a draw in every relative optimum (every valley), but the result will make a large number of agents frustrated. If too many of them feel unhappy, they may decide to continue the voting game, until they climb over the barrier and land in the next valley -- where on the whole they will be just as frustrated as previously, with the only difference that it will not be the same agents that feel frustrated.

\section{Slow dynamics} 
\label{sec_dynamics}

\begin{figure}[H] 
  \centering
  \includegraphics[width=0.7\columnwidth, angle=-90]{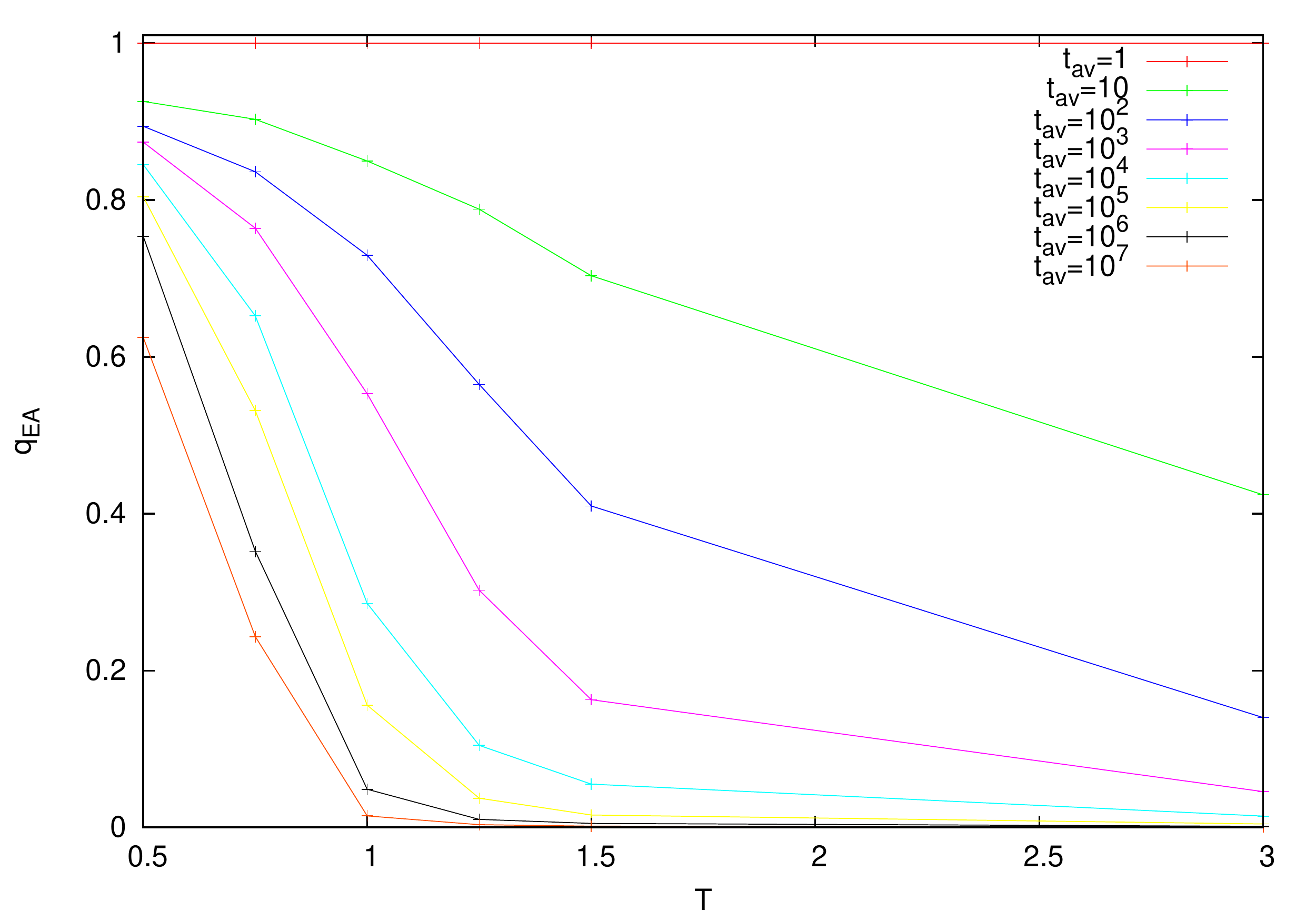}
  \caption{Decay of the Edwards-Anderson order parameter as a function of the noise level (temperature) for a $2d$ system of size $100\times100$. The parameters of the curves are the lengths of the time windows over which the order parameter has been averaged, measured in Monte-Carlo sweeps (1 sweep is about $10^4$ steps).}
 \label{fig_ea_dec}
\end{figure}

We have already mentioned that heterogeneous systems tend to display very slow dynamics. As a demonstration of this kind of evolution, we run extensive simulations on two dimensional samples where the agents were placed on a square lattice of size $100\times100$. The results are shown in Fig.~\ref{fig_ea_dec}. The figure shows the noise dependence of what is called the \citet{13} order parameter in the physics context. This order parameter measures to what extent the agents, exposed to the influence of their peers and that of the noise, preserve the values of their choices $s_{i}$ over time. (The external field has been set to zero in these simulations.) 

\begin{equation}
 q_{EA}=\frac{1}{N}\sum\limits_{i=1}^{N}\left<s_{i}\right>^{2}_{av}
 \label{eq_mmc_2}
\end{equation}	

The averages in Eq.~(\ref{eq_mmc_2}) are taken over various lengths of time, shown as the parameters of the curves. Note that these averaging times are given in units of Monte Carlo sweeps, that is roughly $10 000$ Monte Carlo steps, so the longest runs correspond to $10^{11}$ attempted spin flips. (This is, of course, unnecessarily long in most socioeconomic contexts; we merely display these curves to show the effect of the observation time on the result.) What we can see in Fig.~\ref{fig_ea_dec} is that the order parameter remains rather large even at relatively high noise level (the characteristic scale of noise is fixed by our choice of the coupling strength, that is $\mathcal{O}(1)$), provided the averaging time is not too long. The system appears to be ordered or disordered depending on the length of time period over which we observe it. 

We interpret these results as the manifestations of the rugged phase space landscape: observation for a short period finds the system in a given basin of attraction where a certain ordering pattern prevails and results in a value of $\mathcal{O}(1)$ of the order parameter. A longer observation time allows the system to wander away to conquer larger and larger regions in phase space, and the order parameter will decay accordingly.

Two remarks are in order here: Any structure will gradually disappear in a finite size system, if observed for a long enough period: the system will go round the whole phase space and the traces of order will be erased. What we see here is something different, however. Our system is sufficiently large to make the time period needed to explore the whole phase space astronomical. Just consider that the number of phase space points is $2^{10000}$, so even our longest Monte Carlo runs are negligible in length compared to the number of steps needed to visit every microscopic state. On the other hand, our MC runs are sufficiently long to allow the system to explore larger and larger regions of phase space, explore a low cost, quasi equilibrium region, then escape from it and walk over into another low cost region, etc. Within a given quasi-equilibrium region, that is for a relatively short observation time, the system may seem to be well ordered (in the sense that it preserves its microscopic structure), but watching it over a long period, covering several such quasi-equilibrium regimes, the patterns gradually blur and the order parameter declines. 

The second remark concerns the dimensionality of the system. There is a general agreement in the spin glass community that there is no finite temperature phase transition in a two dimensional spin glass. Our results by no means contradict this belief. A real phase transition would require the building up of macroscopic walls between the phase space valleys as the particle number goes to infinity. We do not see any evidence for this. On the other hand, nothing excludes the presence of finite, even relatively high, barriers in two dimensions which can confine the system for rather long periods. What is perhaps surprising about Fig.~\ref{fig_ea_dec} is the persistence of the order parameter up to rather high noise levels for the shorter observation times. The slow dynamics allows us to measure various observables in wide time windows where the system can be regarded as being in a quasi-equilibrium state.

In the language of the multiagent model, the message of Fig.~\ref{fig_ea_dec} is the following: socioeconomic systems may display shorter or longer periods of quasi stability during which they stay within a relatively small region of their phase space and sustain a relatively stable pattern in their agents' choices, but over much longer periods they will evolve out of this region, the original pattern will decay and a new one will develop. This is reminiscent of the dynamics of the evolution of society or economy.

\section{The network of couplings and the network of the low lying states in phase space} 
\label{sec_phasespace}

Networks are popular today in the most diverse disciplines. Most of the works done on networks tend to focus on their topological properties, although it is widely realized that weights, currents, etc. on the links can make a tremendous difference. All these analyses have the common feature that they explain some property or function of the network on the basis of its local properties: there is this special link that is important because if we remove it the network falls apart; there are these hubs that are important because they are connected to a very large number of other nodes; there are these links that carry most of the traffic; there are those cliques that largely determine the reaction of the whole network, etc. In all these cases, a global property, function, efficiency or stability is explained in terms of local geometrical properties. In this section we would like to emphasize another feature that we consider important: in the artificial society of agents considered in this paper the dynamics is governed by the cost function that is a \textit{functional} defined over the interaction matrix. As a result, the dynamics depends on the whole system of couplings, therefore it may be hard if not impossible to identify that particular player or interaction on which the stability, efficiency, or any other global property of the network really depend. 

In this section, we place our agents on the nodes of a complete graph. This means that every agent directly interacts with every other one. The distribution of (positive or negative) weights on the edges matters, of course. If $N$ is small (less than $20$, say), the spectrum of possible values of the cost and their multiplicities (i.e. the number of different microscopic configurations belonging to the same cost) can be readily determined from Eq.~(\ref{eq_ham_1}) with the help of a computer via complete enumeration (not by Monte Carlo). If all the couplings are $+1$ this does not even require a computer, but, as indicated before, we wish to consider samples with an approximately equal number of positive and negative couplings. The pictures of dense networks with even a moderate number of nodes can become totally confusing, so for the sake of illustration of a few generic points, we will be considering really small systems here, with $N$'s as small as $6$ or $16$. We will also set the external field to zero, for simplicity.

\begin{figure}[H] 
  \centering
	\begin{tabular}{cc}
  \includegraphics[width=0.45\columnwidth]{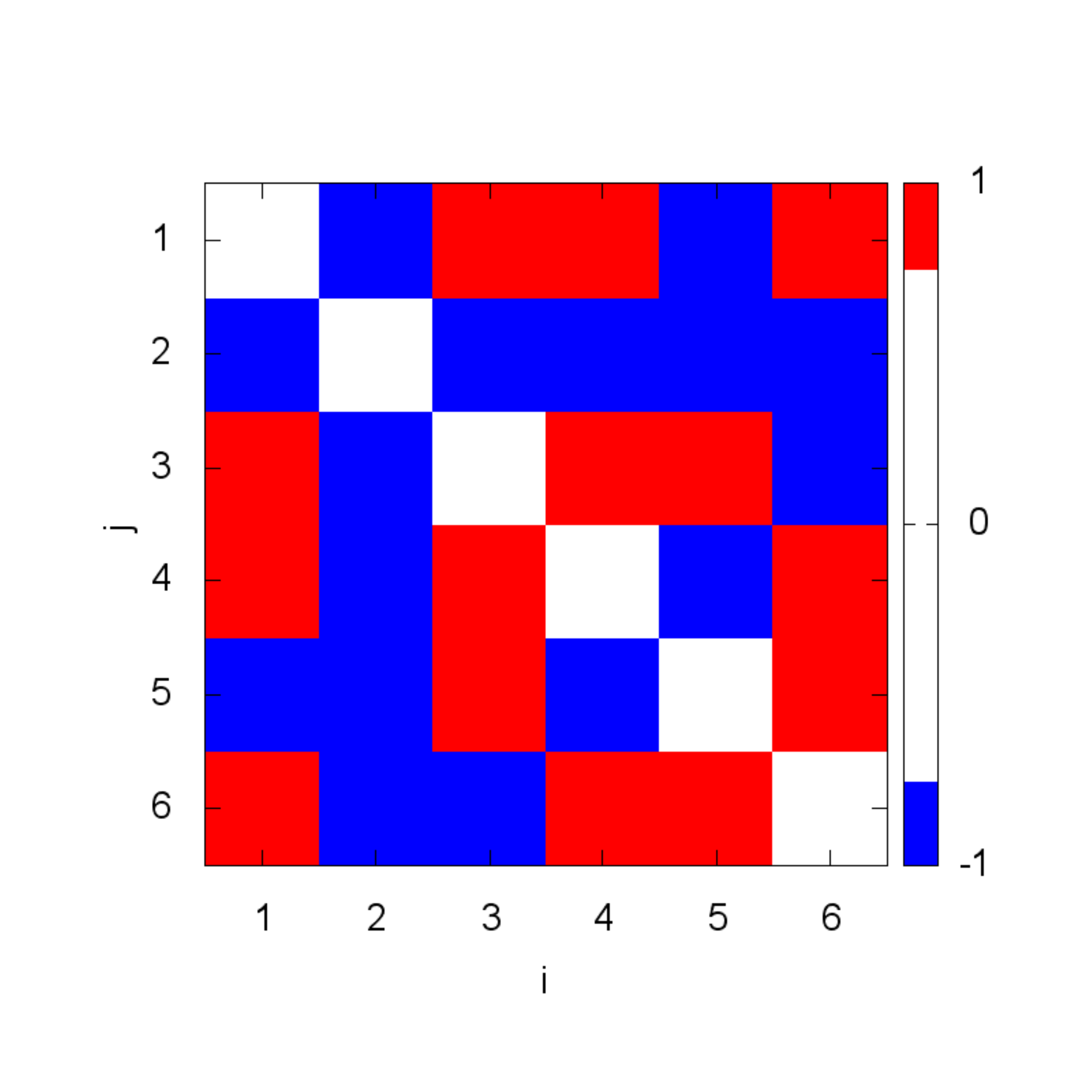}&
	\includegraphics[width=0.45\columnwidth]{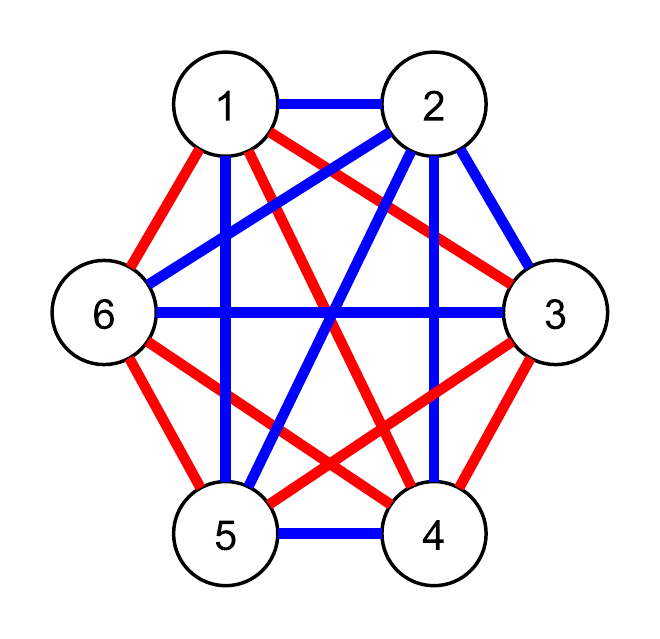}
	\end{tabular}
  \caption{The interaction matrix and network of couplings for a given small sample.}
 \label{fig_adj_1}
\end{figure}

Fig.~\ref{fig_adj_1} shows the matrix of couplings (interaction matrix) for such a small system. The positive ($+1$) couplings are coloured red, the negative ($-1$) ones blue. The corresponding network is also displayed in the figure.

\begin{figure}[H] 
  \centering
	\begin{tabular}{cc}
  \includegraphics[width=0.3\columnwidth]{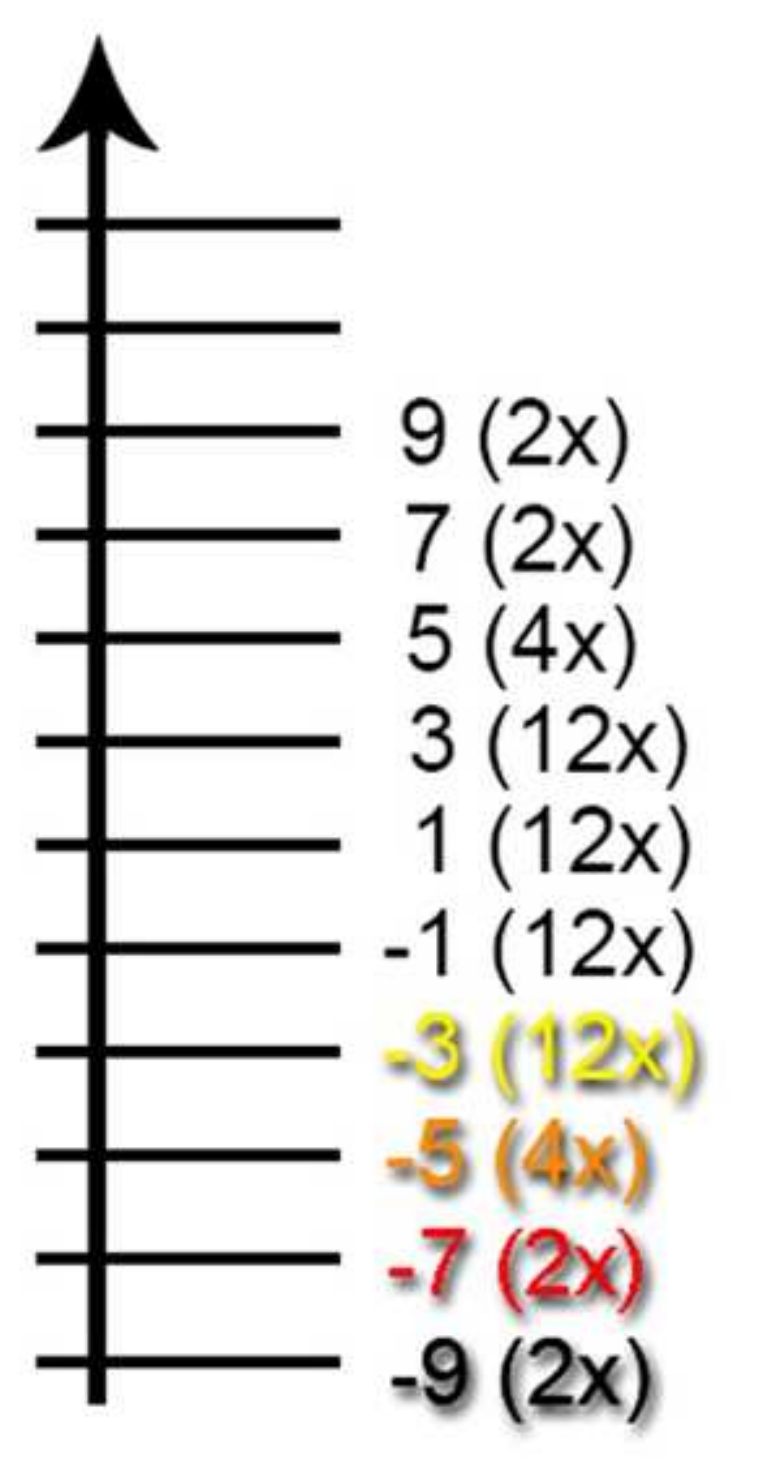}&
	\includegraphics[width=0.6\columnwidth]{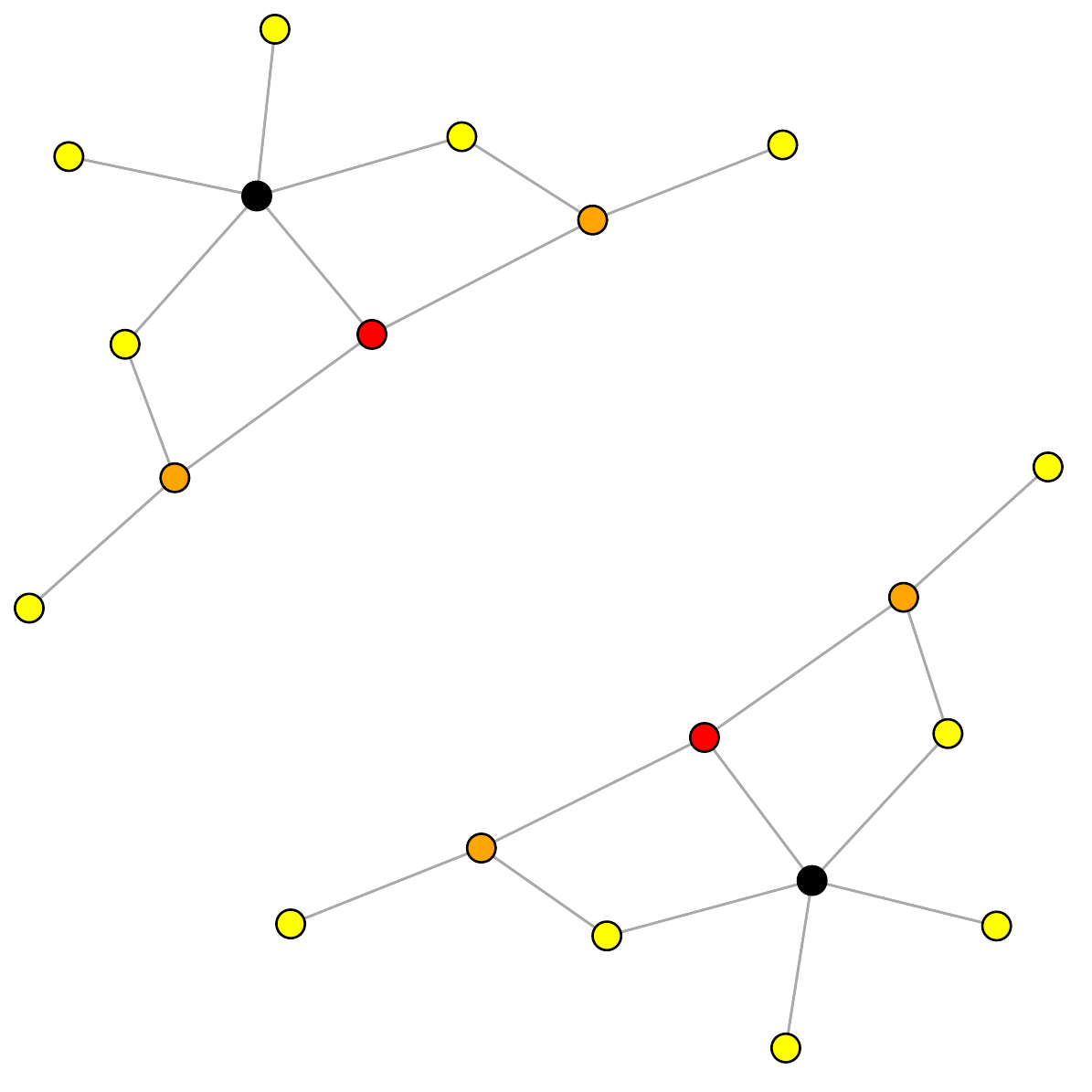}
	\end{tabular}
  \caption{Cost spectrum and the network of the four lowest cost states for the system shown in Fig.~\ref{fig_adj_1}. (Complete enumeration in phase space.)}
 \label{fig_ps_1}
\end{figure}

The spectrum of this system (i.e. the set of values of the cost function in the different microscopic states) is given in Fig.~\ref{fig_ps_1}. There are $2^N = 64$ microscopic states, but the number of different costs (for this particular distribution of couplings) is only $10$. Of these, the phase space network of the lowest $4$ cost states is also shown in the figure. The two lowest cost states are the black nodes, the next levels are the two red ones, the next are the $4$ orange nodes, and the $12$ yellow nodes are the fourth level states.

In the figure we connected those states that can be reached from each other by flipping a single agent. As can be seen, the lowest lying states are divided into two groups: any path crossing over between them should climb through states with a higher cost than the ones shown in the figure. (Of course, if we displayed all the nodes and links there would be no such division between the states.) 

This way we have defined a network in the space of the microstates, the network of low cost states. This network is very different from the network of couplings, and depends functionally on it: the structure of the phase space network depends on the complete set of couplings and their concrete distribution over the agents.

Now we can set our mini-society in motion, according to the Monte Carlo dynamics. No matter in which microstate do we start the simulation, at low noise levels it will quickly locate the lowest lying microstates, one of those depicted in Fig.~\ref{fig_ps_1}. Hereafter  the system is executing a random walk on the phase space network, in principle not only on the portion shown in Fig.~\ref{fig_ps_1}, but also over the whole set of microstates. However, at low noise levels only the lowest cost states will be visited, it is highly improbable that the repeated application of the update rule Eq.~(\ref{eq_mmc})  would produce such a series of consecutive bad moves that would drive the simulation up to higher lying states, thereby allowing the simulation to traverse the cost barrier to the other set of low lying states. This way, at a given level of noise and within a given time window (a given length of Monte Carlo run) the two halves of the phase space become mutually inaccessible from each other under the rule Eq.~(\ref{eq_mmc}). The system gets frozen in one or the other half of the phase space. 
\begin{figure}[H] 
  \centering
	\begin{tabular}{cc}
  \includegraphics[width=0.45\columnwidth]{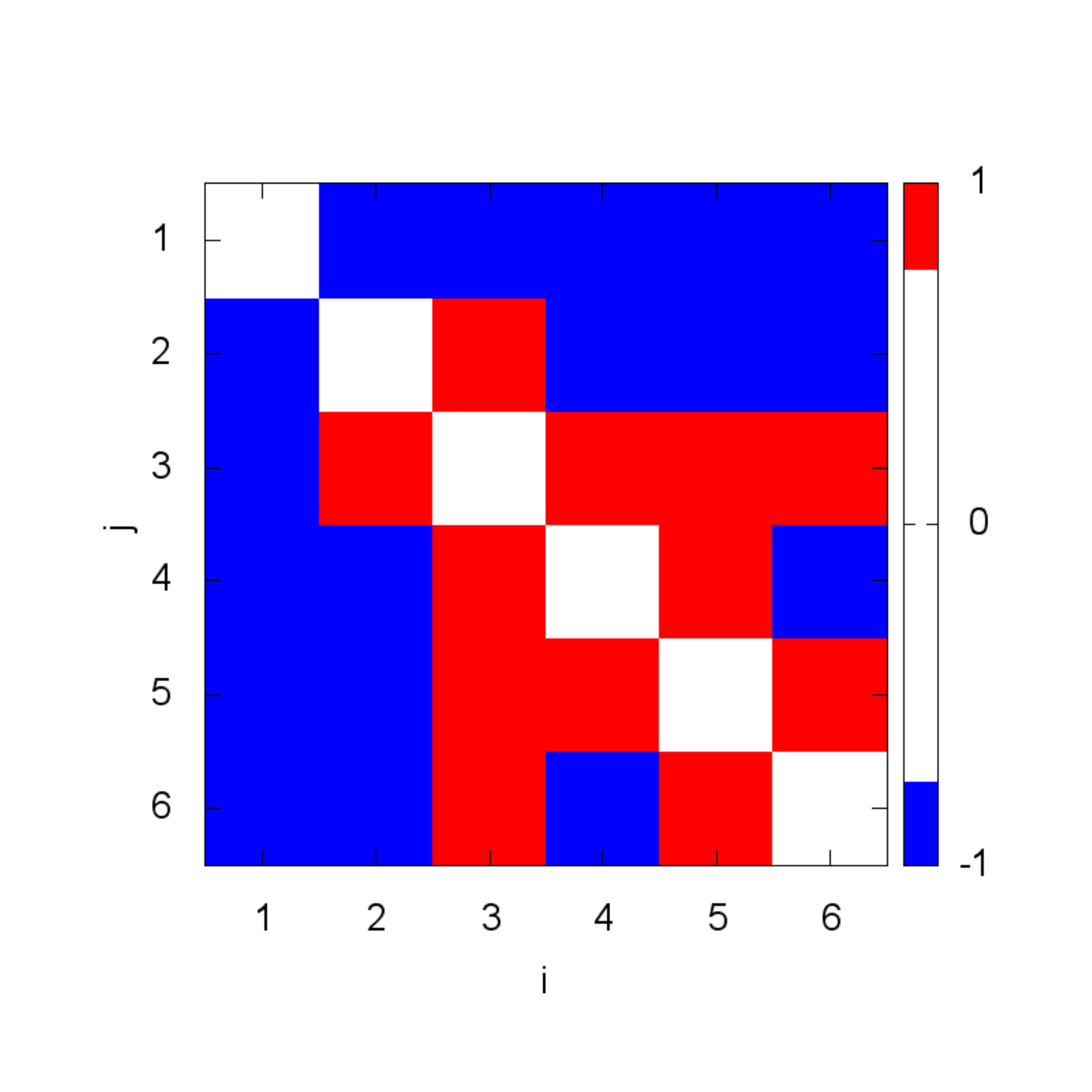}&
	\includegraphics[width=0.45\columnwidth]{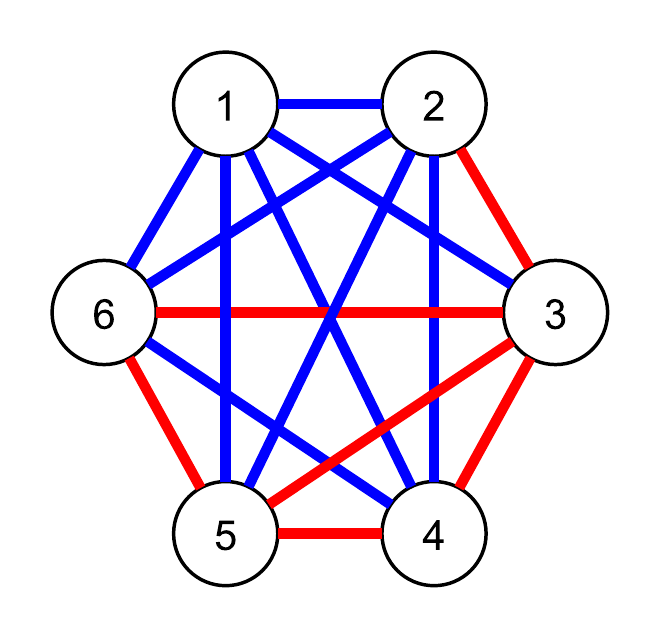}
	\end{tabular}
  \caption{The interaction matrix and network of couplings for a sample seemingly very different from the previous one, yet having the same spectrum and phase space landscape.}
 \label{fig_adj_2}
\end{figure}

Fig.~\ref{fig_adj_2} displays a seemingly very different arrangement of couplings. It turns out, however, that the cost spectrum and the phase space structure of this system are exactly the same as those belonging to the previous interaction matrix. This is related to an exact symmetry, the gauge symmetry of the model, \citet{15}, and is due to the simple structure of the interaction considered here.

\begin{figure}[H] 
  \centering
	\begin{tabular}{cc}
  \includegraphics[width=0.45\columnwidth]{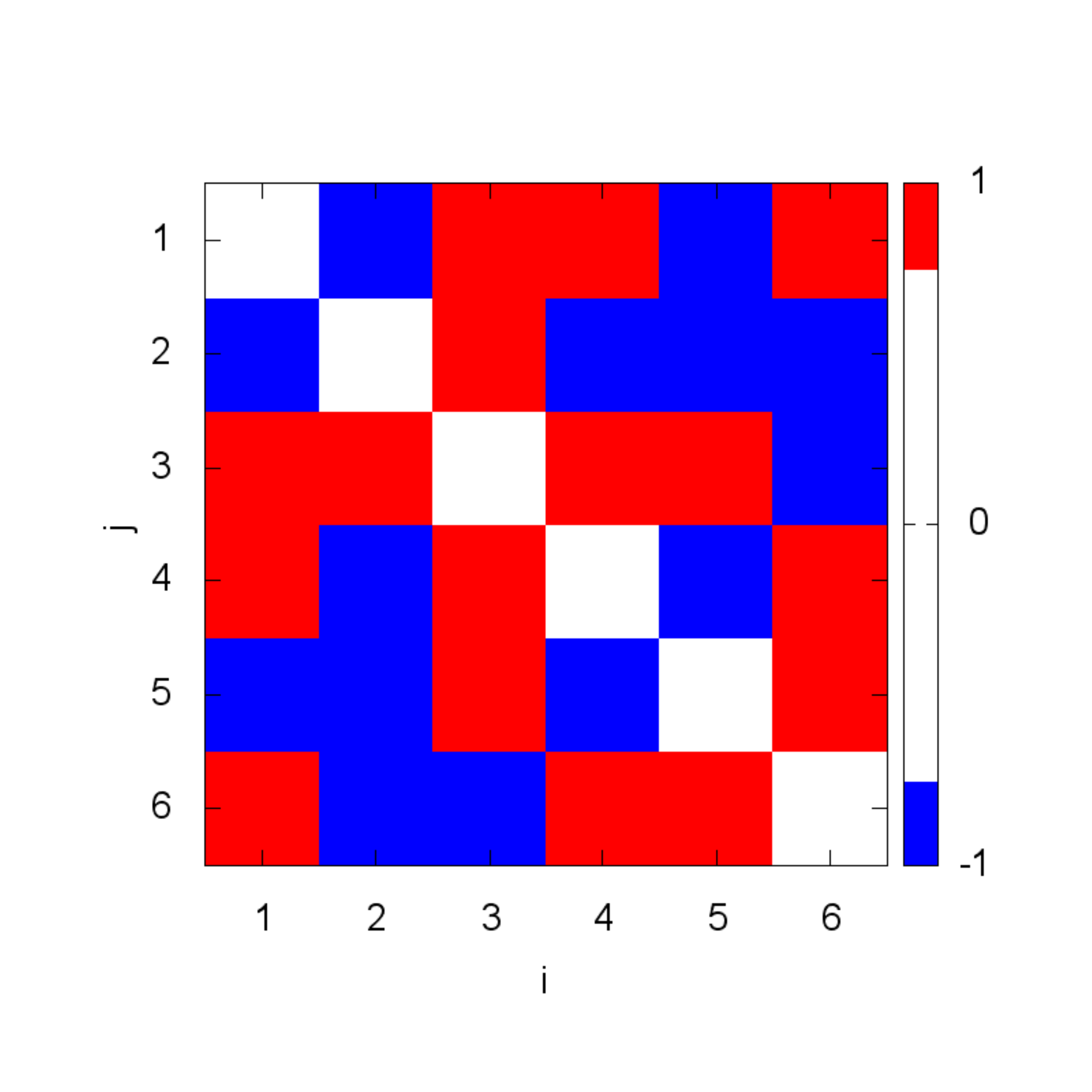}&
	\includegraphics[width=0.45\columnwidth]{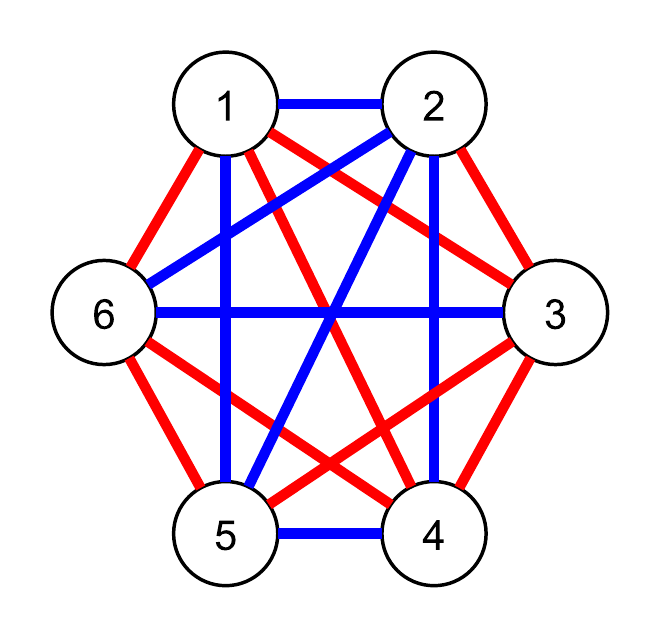}
	\end{tabular}
  \caption{The interaction matrix and network of couplings for a sample that is different from the one in Fig.~\ref{fig_adj_1} only in the sign of a single coupling.}
 \label{fig_adj_3}
\end{figure}

\begin{figure}[H] 
  \centering
	\begin{tabular}{cc}
  \includegraphics[width=0.3\columnwidth]{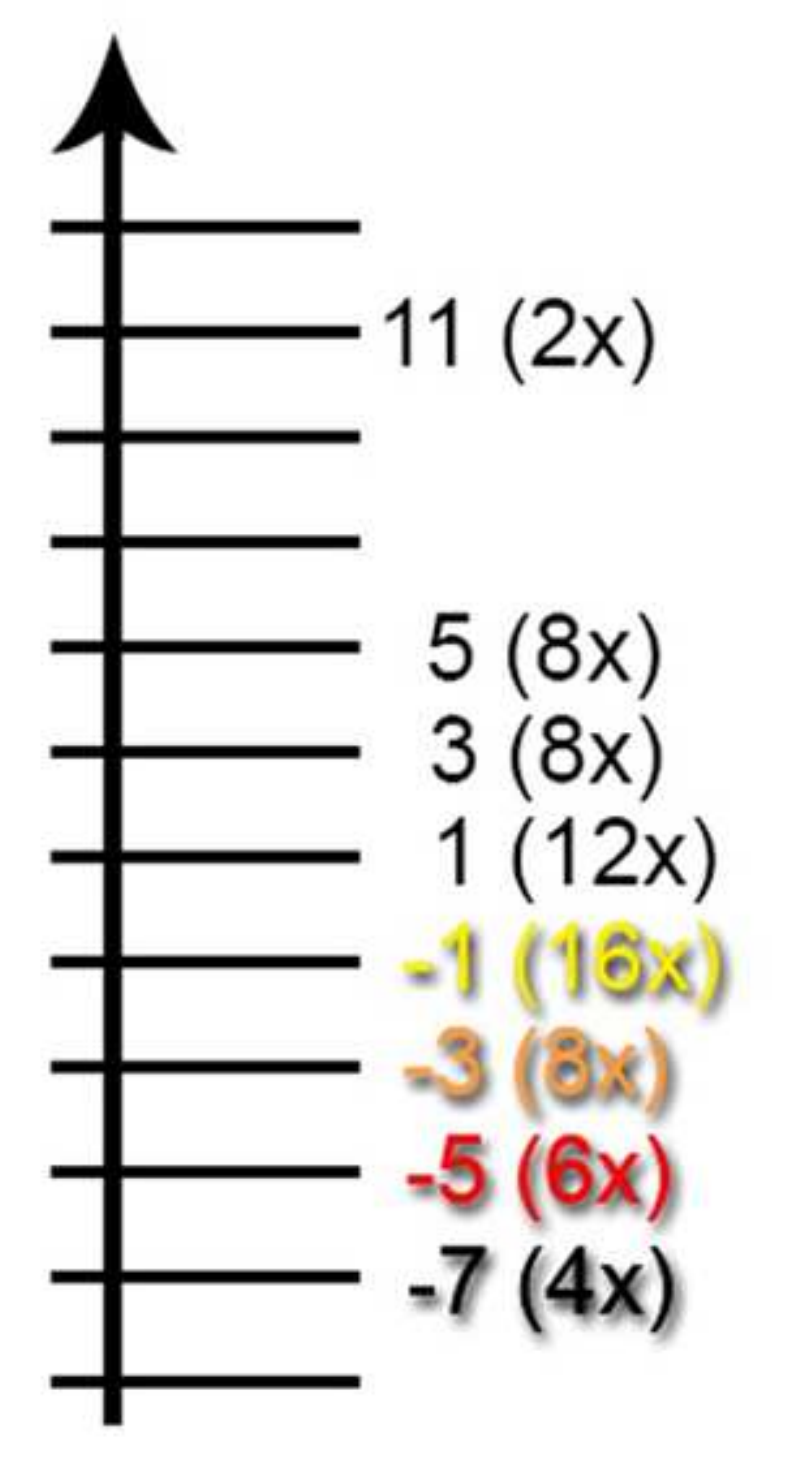}&
	\includegraphics[width=0.6\columnwidth]{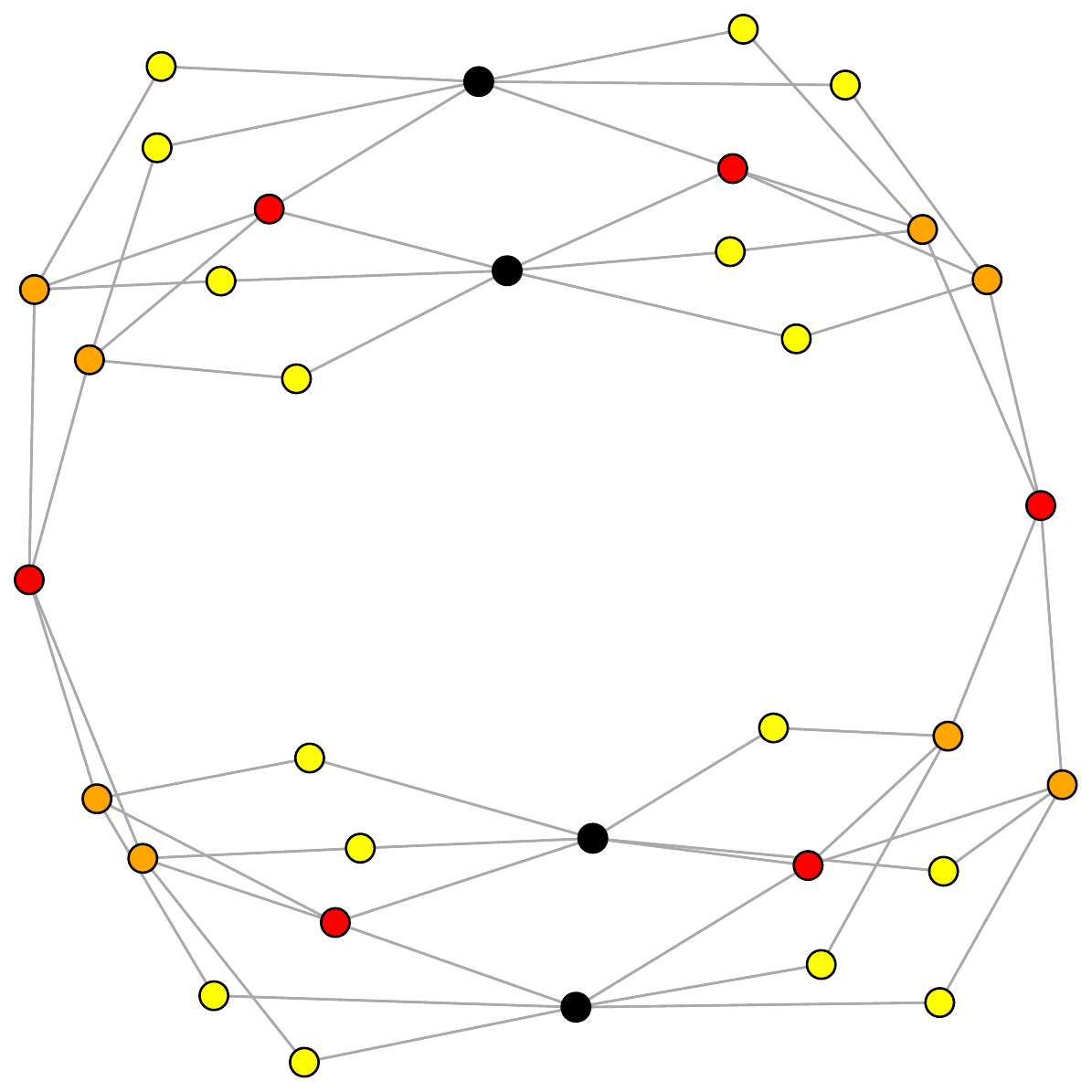}
	\end{tabular}
  \caption{Cost spectrum and the network of the four lowest cost states for the system shown in Fig.~\ref{fig_adj_3}. Note that the change of a single sign has rearranged the spectrum and the landscape. (Complete enumeration of phase space.)}
 \label{fig_ps_3}
\end{figure}

The effect illustrated in Fig.~\ref{fig_adj_3} is the opposite of the above: here we are considering an interaction matrix that differs from the one in Fig.~\ref{fig_adj_1} only in the sign of a single coupling, yet the resulting spectrum and the structure of the phase space landscape, displayed in Fig.~\ref{fig_ps_3}, are rather different from those shown in Fig.~\ref{fig_ps_1}.

We can see in Fig.~\ref{fig_ps_3} that the number of ground states is now $4$, and they are all mutually accessible from each other via routes that go through states having not higher than the third cost level.
 
This is an illustration of the sensitivity of the dynamics to small details. The small systems shown here are, of course, just the simplest illustrations of this feature: we chose them so small to keep the figures comprehensible. We shall look at the same phenomenon from different angles in the following.

\begin{figure}[H] 
  \centering
  \includegraphics[width=\columnwidth]{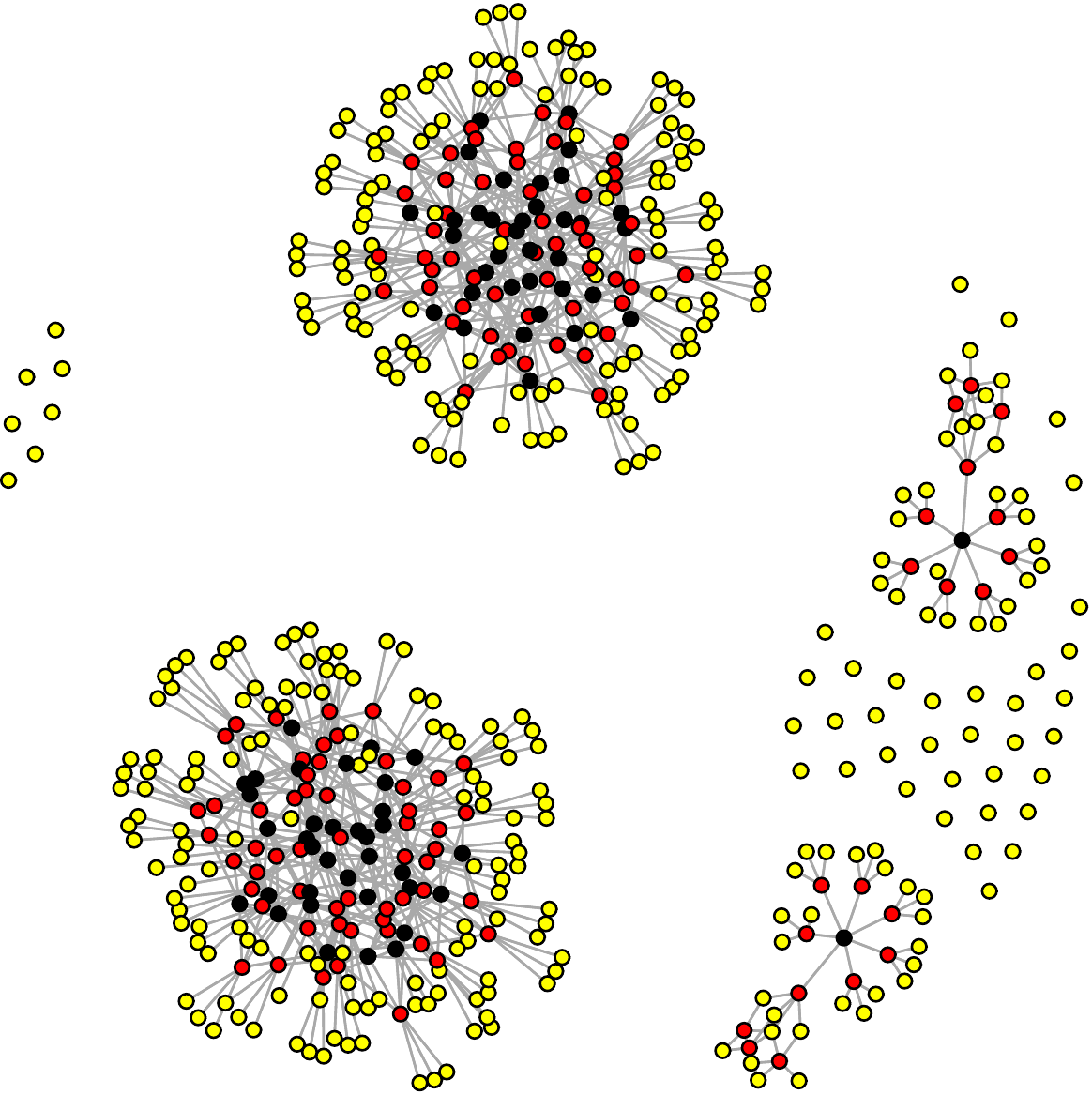}
  \caption{Phase space landscape for a system of $N=16$ agents on the complete graph. Only those of the lowest $3$ cost states are shown that can be reached from each other at a noise level $T=0.2$ with a probability larger than $1\%$. Note that the phase space network falls apart into distinct clusters that are hard to reach from each other. (Complete enumeration of phase space.)}
 \label{fig_sp_split}
\end{figure}

The network shown in Fig.~\ref{fig_sp_split} belongs to a system of size $N=16$. Only the lowest $3$ cost states are shown, as a matter of fact not even all of them, only those that can be reached from each other with a probability above $1\%$ at the given noise level of $T=0.2$. It is clear that the structure of phase space network quickly becomes very complicated as the number of agents increases. The network is now divided into several valleys formed by states that are very strongly connected among themselves, but nearly isolated from the rest of the phase space. (Some of these valleys consist of a single state.). The dynamics of the system will crucially depend on where the walk is started. Once the system gets trapped in one of the valleys, it will spend a very long time there before it can escape.

As we raise the noise level, more and more valleys will coalesce, i.e. become accessible from each other. Given the exponential dependence of the acceptance probability Eq.~(\ref{eq_mmc}) on noise, a very small shift in the value of $T$ is enough to melt many valleys together. We can see then that we are dealing with a very bad case of ergodicity violation here: not only do we have a large number of valleys, but even the landscape is very sensitive to any change in the distribution of couplings and in the control parameters. We believe that whenever a system encompasses the elements of competition and collaboration a similar violation of ergodicity will result.

Another interesting aspect of these phase space networks is the following. Suppose we observe the system as it is performing its random walk inside a strongly connected cluster of states, such as those indicated in Fig.~\ref{fig_sp_split}. After a certain time it finds an escape route into the next valley. There may be several such paths, some climbing over higher passes, some going through lower ones. It may happen that the number of high passes is large, while the low passes are so rare that the probability of a random walker finding them is nearly zero. Alternatively, it may be that the paths through the low passes are very long, so walking along them would require a highly improbable series of random steps. Whichever should be the case, it will lead to a situation where escaping along relatively high cost paths may turn out to be more probable than along a low cost one, if the former are more numerous, or more expedient for any other reason. In addition, the balance between the two options will depend again very sensitively on the level of noise present in the system. Without noise the system would never escape from any valley, however shallow.

\section{The distribution of correlations on the complete graph} 
\label{sec_correlations}

First, let us define the quantities we wish to discuss in this section. Following the use of terminology in physics, we will call the quantity
\begin{equation} 
K_{i,j}=\left<s_is_j\right>
 \label{eq_K}
\end{equation}
the correlation between agents $i$ and $j$, while the quantity
\begin{equation} 
C_{i,j}=\left<s_is_j\right>-\left<s_i\right>\left<s_j\right>
 \label{eq_C}
\end{equation}
will be referred to as the connected correlation. 

The averages denoted by $\left<\dots\right>$ in the above formulae are expectation values calculated with the Boltzmann weight 

\begin{equation} 
\frac{1}{Z}e^{-\frac{E}{T}}
 \label{eq_boltzmann}
\end{equation}
($E$ is the cost and $Z$ is a normalization factor), when a full enumeration is possible; or along a suitably long Monte Carlo trajectory, when the system's size does not allow full enumeration.

In a heterogeneous system like a spin glass these correlations depend on the indices in an essentially random fashion for each sample. For that reason, they have not been considered legitimate objects of study in most of the physics literature. (Among the few exceptions the paper \citet{16} is of special significance for us.) Instead of Eq.~(\ref{eq_K}) or Eq.~(\ref{eq_C}), the spin glass literature concentrated on well-behaved, smooth and self-averaging combinations of the squares of these quantities, \citet{12}, \citet{17}. We will keep to the simple correlations as defined above. 

In this section we consider the case of the complete graph again, and determine the $N(N-1)/2$ correlation values either by full enumeration or by Monte Carlo. These objects will depend on the pair $i,j$, on the noise and the external field (if there is one). Imagine we arrange these values in the form of a histogram. According to the definition Eq.~(\ref{eq_K}), the support of this distribution will be the interval $[-1,+1]$. At high levels of noise all the correlations will be very small, that is we will have a sharp peak at the origin and essentially zero elsewhere. As the noise level is lowered, more and more correlations acquire larger (positive or negative) values, the peak at the origin broadens. The broadening of the distribution of correlations will be the central issue in the following discussion.

\begin{figure}[H] 
  \centering
  \includegraphics[width=\columnwidth]{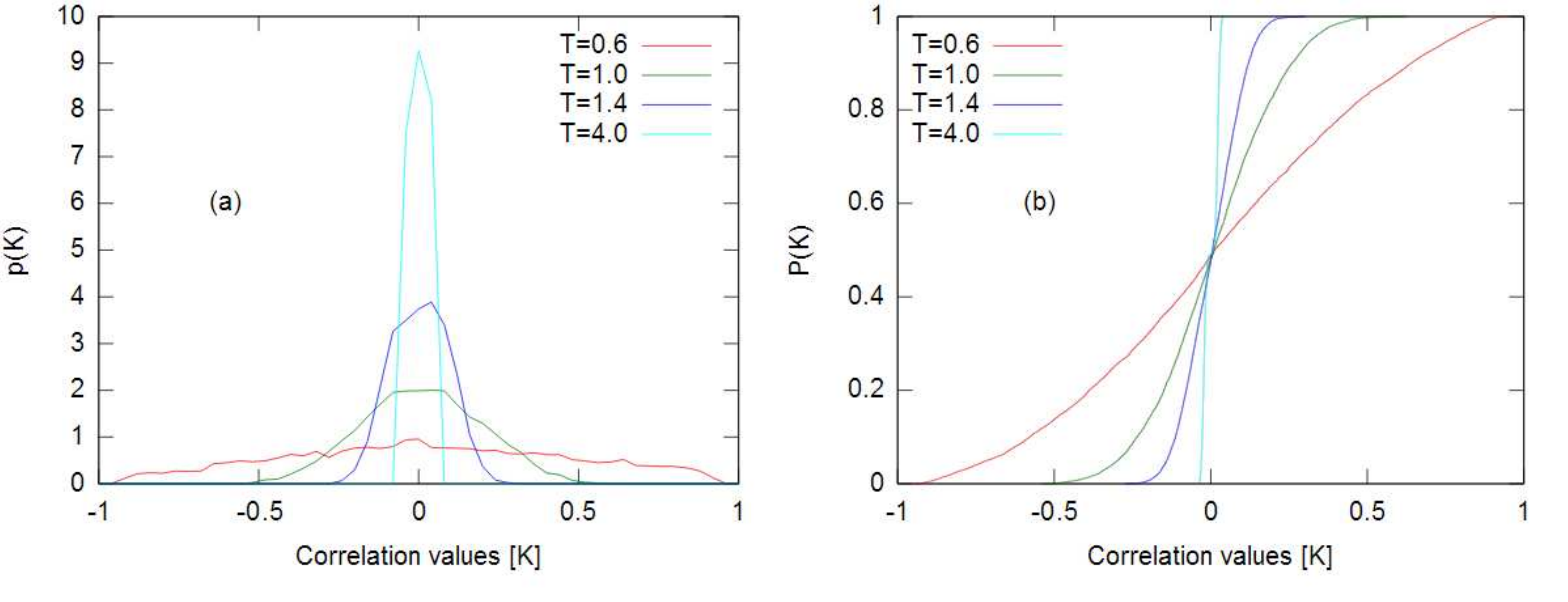}
  \caption{($a$): Probability-density of correlations for a system of size $N=128$, for a single sample on the complete graph at different noise levels. (Metropolis Monte-Carlo simulations, run time: $10^7$ spin flops; $h = 0$) ($b$): Cumulative distribution functions for the same system. Note that the fluctuations about the smooth curves are so small that they are masked by the width of the curves.}
 \label{fig_hist_base}
\end{figure}

Let us look at the histogram or probability density of correlations, obtained by Monte Carlo simulations on an $N=128$ system at different noise levels, as given in Fig.~\ref{fig_hist_base}($a$). The broadening of the central peak with decreasing noise is clearly visible; at $T=0.6$ we have an almost completely flat distribution. The behaviour of the cumulative distribution corresponding to the density in Fig.~\ref{fig_hist_base}($a$) is shown in Fig.~\ref{fig_hist_base}($b$). 

\begin{figure}[H] 
  \centering
  \includegraphics[width=\columnwidth]{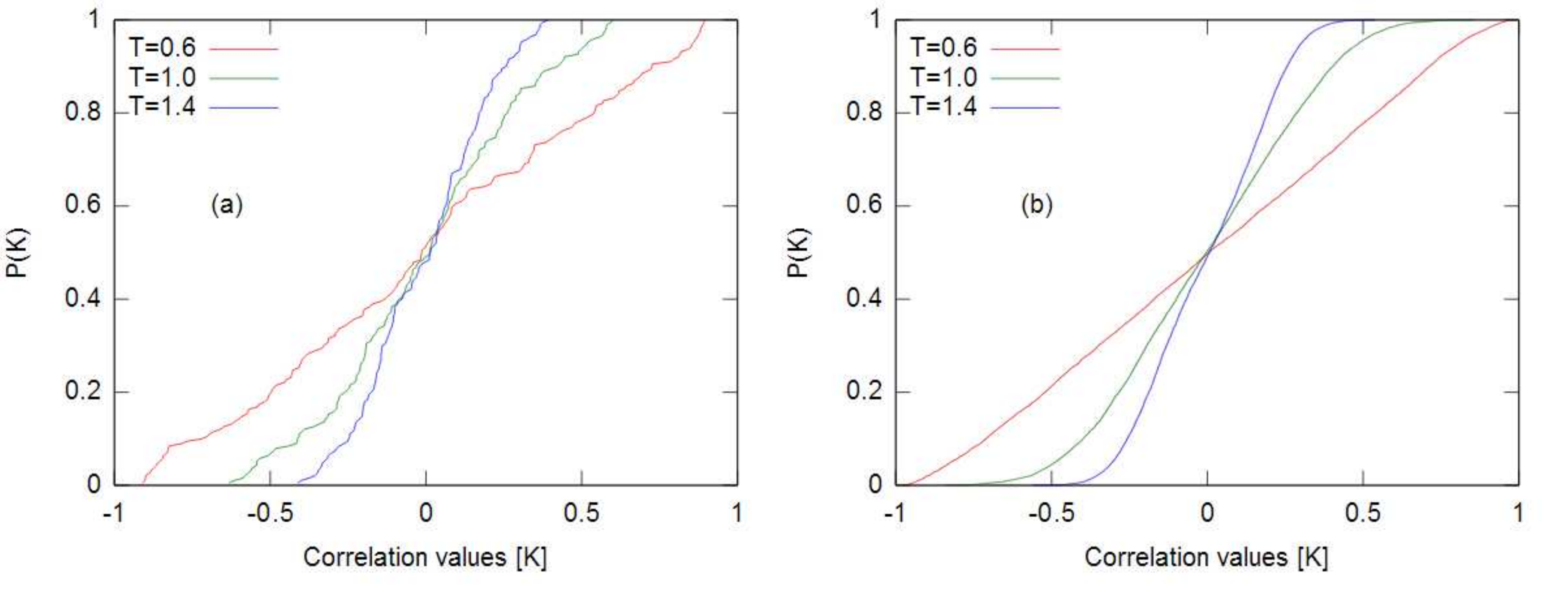}
  \caption{($a$): Cumulative distribution-function for a system of size $N=20$, for a single sample on the complete graph at different noise levels. (Complete enumeration in phase space.)
($b$): Cumulative distribution-functions averaged over $50$ samples of size $N=20$, on the complete graph at different noise levels. (Complete enumeration in phase space.)}
 \label{fig_hist_avg}
\end{figure}

\begin{figure}[H] 
  \centering
  \includegraphics[width=\columnwidth]{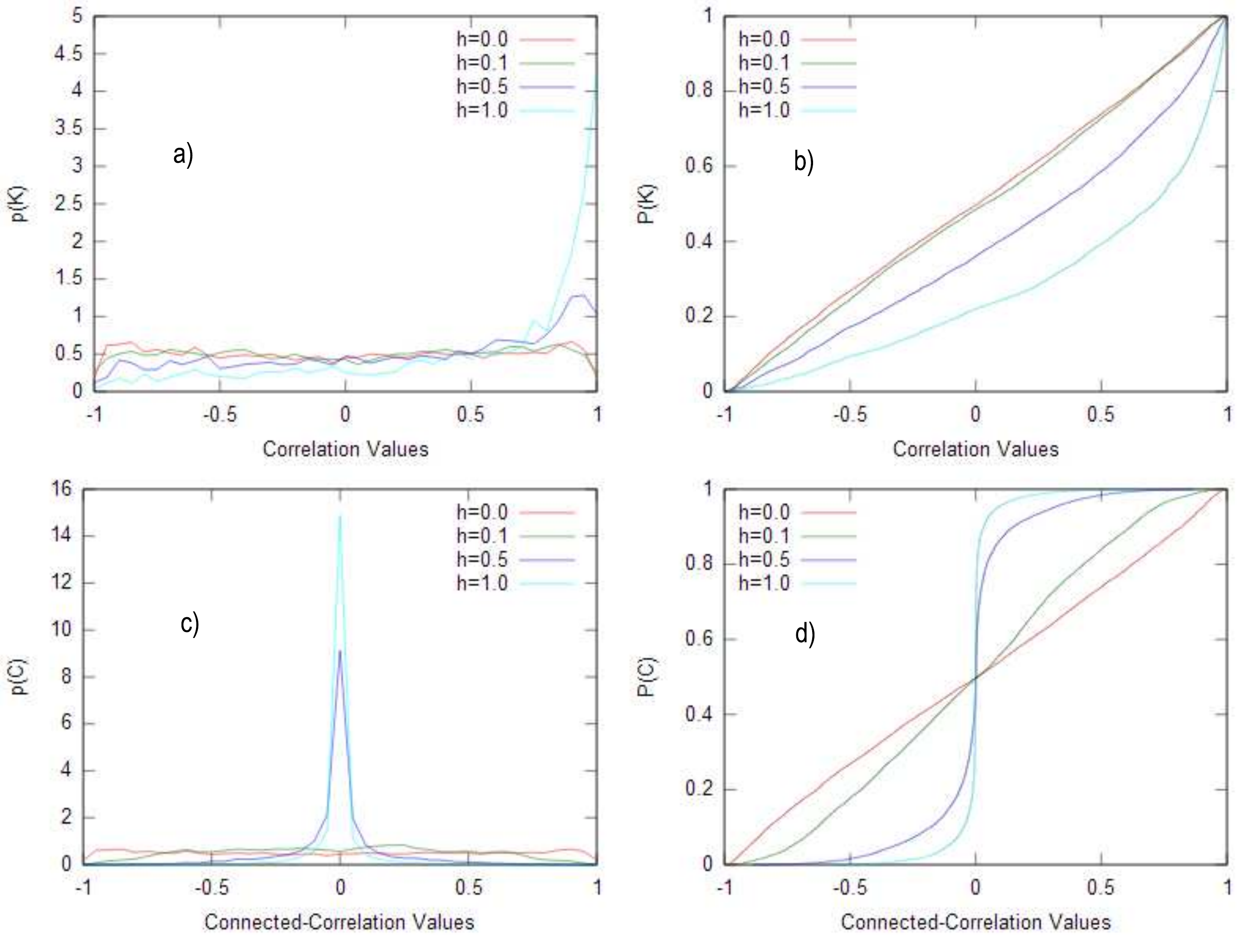}
  \caption{($a$): Probability-density of correlations for systems of size $N=20$, averaged over $50$ samples on the complete graph at different external fields at noise level $T=0.6$. (Complete enumeration.) ($b$): Cumulative distribution-functions of correlations for the same system as in ($a$). ($c$): Probability-density of connected correlations for the same system as in ($a$). ($d$): Cumulative distribution-functions of connected-correlations for the same system as in ($a$).}
 \label{fig_hist_h}
\end{figure}

\begin{figure}[H] 
  \centering
  \includegraphics[width=\columnwidth]{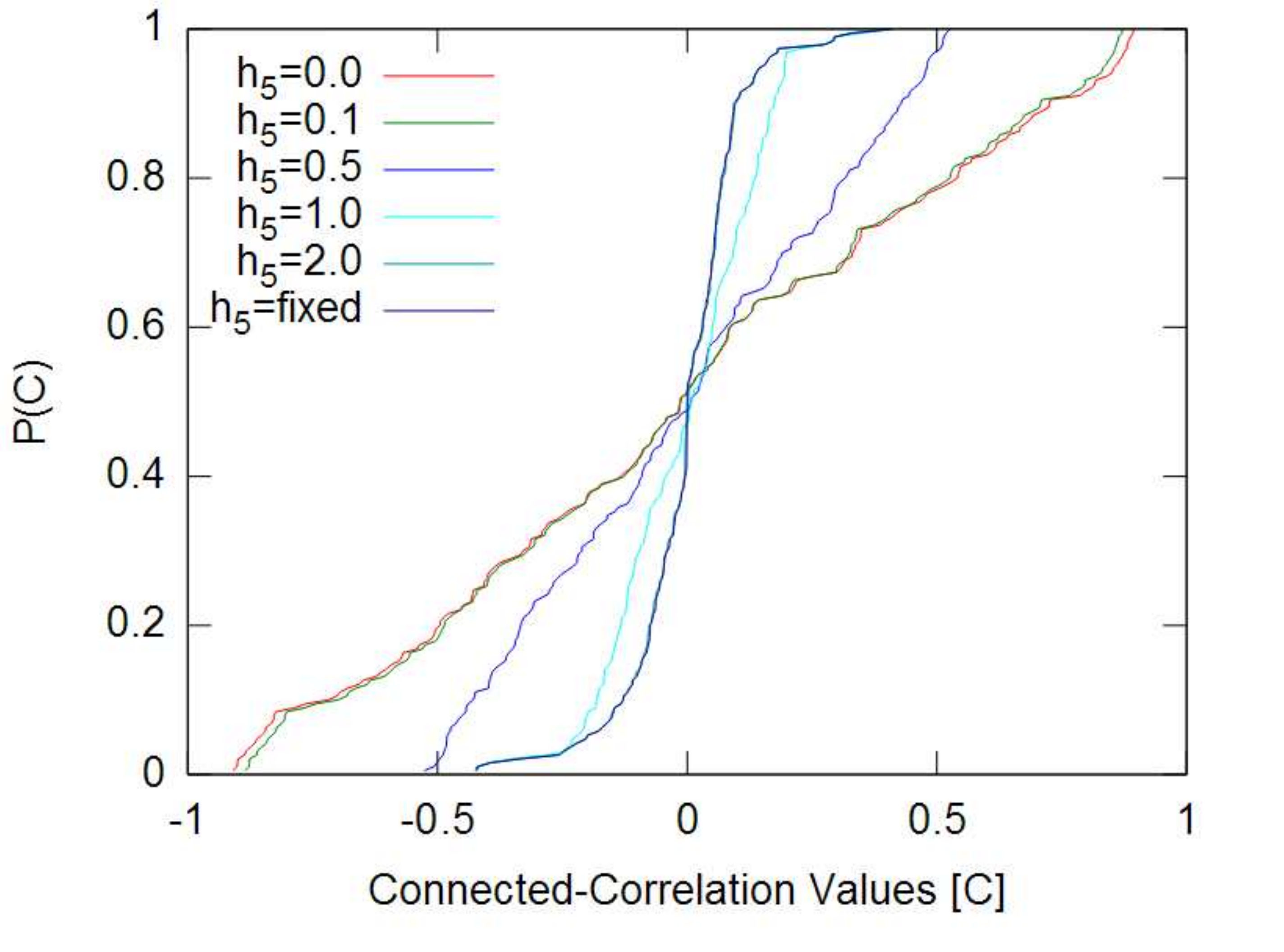}
  \caption{Cumulative distribution of connected correlations in a single sample on a complete graph of size $N=20$, at noise level $T=0.6$. The curves correspond to local fields of different strengths acting on a \textit{single agent belonging to the strongly correlated cluster}, up to a situation when the agent is fixed (as if an infinitely strong field acted on it). The effect of a local field acting on a single strongly correlated agent is similar to the effect of a homogeneous field acting on all of them. (Complete enumeration in phase space.)}
 \label{fig_hist_strong}
\end{figure}

\begin{figure}[H] 
  \centering
  \includegraphics[width=\columnwidth]{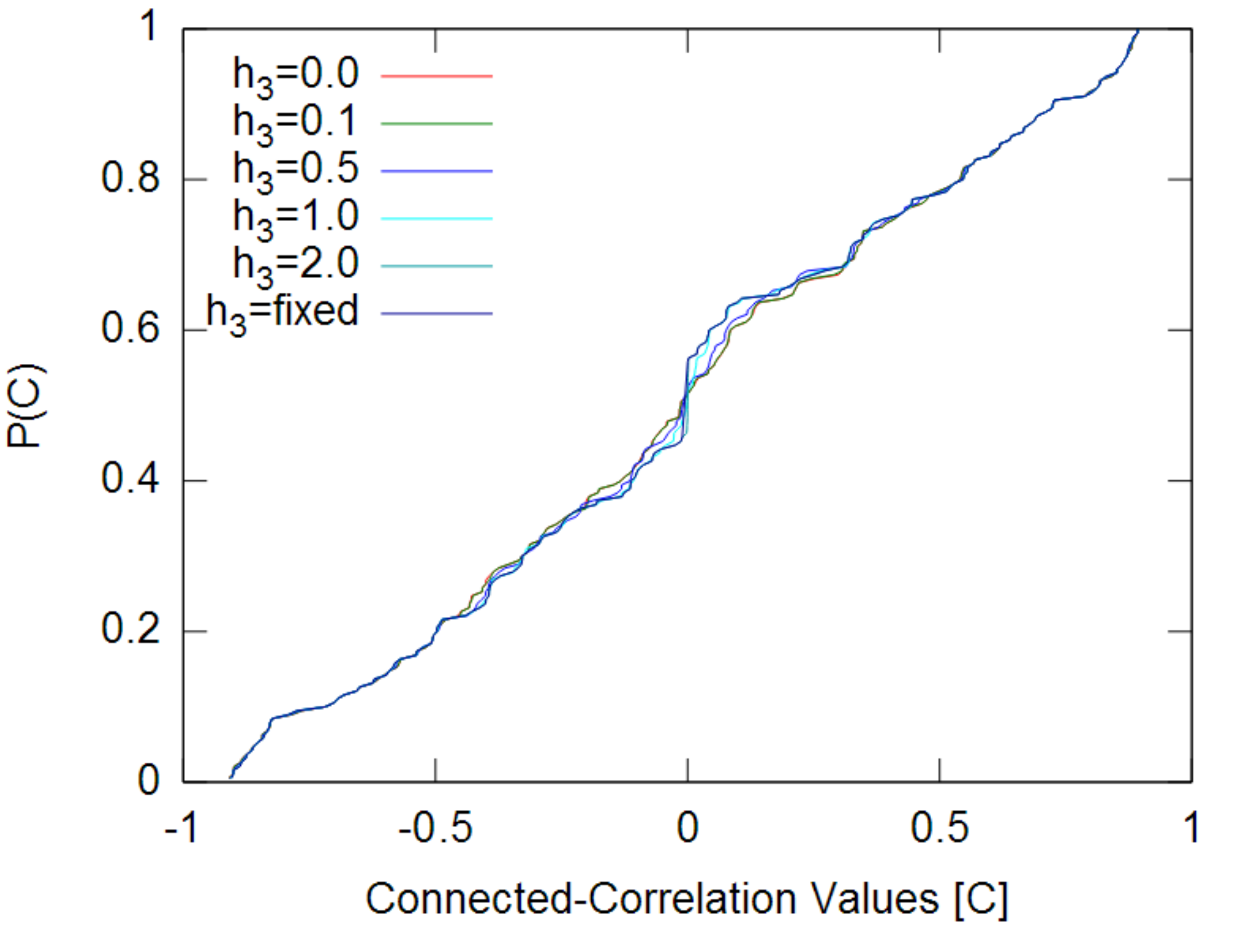}
  \caption{Cumulative distribution of connected correlations in a single sample on a complete graph of size $N=20$, at noise level $T=0.6$. The curves correspond to local fields of different strength acting on a \textit{single agent outside the strongly correlated cluster}, up to the point where the agent is fixed. Such a local field has virtually no effect on the distribution of correlations. (Complete enumeration in phase space.)}
 \label{fig_hist_weak}
\end{figure}

These and the results shown in Fig.~\ref{fig_hist_avg} below are in full agreement with the findings of \citet{16}. We decided to repeat some of the small size measurements reported in that paper, in order to have a solid platform of comparison with our new results in an external field, a situation not covered in there. The finite field results are shown in Figures \ref{fig_hist_h} - \ref{fig_hist_weak}. Furthermore, in order to have a benchmark to which MC measurements can be compared, we have performed the calculation of correlations and connected correlations also by complete enumeration. This, of course, limits the accessible sizes, but apart from some jittering of the curves due to the small size, the results shown in the next two figures display the same overall features as the simulations on larger systems.

The sigmoid-like curves in Fig.~\ref{fig_hist_avg}($a$) show the cumulative distribution function corresponding to the histogram of correlations at different noise levels. This figure belongs to a given random sample of size $N=20$ and has been obtained by complete enumeration. The same curves for a different sample would look roughly like the ones displayed. Averaging over a set of $50$ samples is enough to smooth out the sample dependence in these curves; this is shown in Fig.~\ref{fig_hist_avg}($b$).

The relatively steep rise of the blue curve at $T=1.4$ in Fig.~\ref{fig_hist_avg} corresponds to a concentration of the correlation values around zero. As we go into the low noise region (red curve), this concentration disappears completely, the cumulative distribution approaches a straight line, corresponding to a uniform density over the whole support $[-1,+1]$. This striking effect, already reported in \citet{16}, means that at low noise levels the heterogeneous interactions between the agents result in a distribution of correlations in which any correlation value appears with equal probability. This behaviour is characteristic of the samples with a more or less balanced distribution of positive and negative couplings. For a sample with uniformly positive couplings the distribution would not broaden, but would remain sharp and would merely shift away from the origin.

The connected correlations would not be any different in this setup: the averages $\left<s_i\right>$ are identically zero if we are taking the average over the whole phase space in zero external field.

The situation is quite different if an external field is applied. In Fig.~\ref{fig_hist_h} we present the results obtained by complete enumeration for the distributions resp. cumulative distributions of both the correlations and the connected correlations as functions of the external field, for a system of size $N=20$ at a low noise level ($T=0.6$). At this low noise level and without an external field, the distribution of correlations would be quite flat (see the red curve in Fig.~\ref{fig_hist_h}($a$)). As the field strength increases, more and more of the weight of the distribution gets shifted to the right: the external pressure forces all the agents to choose $s_i = +1$ (purple and blue curves in Fig.~\ref{fig_hist_h}($a$)). Fig.~\ref{fig_hist_h}($b$) tells the same story in the language of the cumulative distribution. 

Now the connected correlation function displays a different behaviour. The external field polarizes the agents, $\left<s_i\right>$ is not zero any more. From a flat distribution valid at zero or small fields (red and green curves in Fig.~\ref{fig_hist_h}($c$)) the density evolves into a sharp peak at the origin (purple and blue curves) as the field increases. Fig.~\ref{fig_hist_h}($d$) is the corresponding cumulative distribution.

 The summary of what we have learned is the following: the distribution of connected correlations strongly broadens as we go deeper and deeper into the low noise region in zero field, until it becomes entirely flat. If we now apply an external field at this low noise level, the distribution shrinks back again into a sharp peak as the field increases.

The beginning and the end of this process is easy to understand. In terms of e.g. the voting interpretation of the model, the description of the situation is this: if everybody goes to the polls totally drunk (high noise level) there will be no correlation between their votes, the distribution of correlations will be sharply peaked at zero. If all of them vote under the strong influence of, say, political intimidation instead of alcohol, they will cast their votes more or less uniformly, so there will be a strong correlation between them, and these correlations will be sharply peaked again, this time centred on a nonzero value. The nontrivial part of the story is what happens in between. When there is not much noise and no external pressure, so agents vote only under the influence of their peers, the distribution of correlations between these votes will be very broad, essentially uniform, even if the number of plus and minus votes is the same. In the physics context, the broadening of the distribution of correlations with decreasing noise level, and its shrinking back again in strong fields is related to the problem of phase transition in an external field in spin glasses, but we do not wish to analyse this link any further here.  

The existence of strong correlations in the interaction-dominated, low noise - low field region does not mean that every single agent is subject to an equally strong pressure from her partners. There always will be some on whom the contradictory influences from their peers nearly or exactly cancel. These will effectively be decoupled from the rest and if they get subjected to an external impact, they will not share this with their partners. If, however, this external factor acts on a strongly connected agent, its effect will be felt across the whole system. Such a focused external impact can be modelled by an external field that targets only a single agent. The illustrations of its effect are exhibited in Figures \ref{fig_hist_strong} and \ref{fig_hist_weak}.

Fig.~\ref{fig_hist_strong} shows the cumulative distribution of connected correlations in a given, $N=20$ sample at noise level $T=0.6$ with an external field of varying strength acting on a \textit{single agent that belongs to the strongly correlated cluster}. Fixing this particular agent is as if a very strong local field acted upon it; this is also shown in the figure. The curves were obtained by complete enumeration. Apart from their jittery character, they are quite similar to those that we obtained when the system was subjected to a \textit{uniform field}, Fig.~\ref{fig_hist_h}($d$). An impact on a member of the strongly correlated cluster has the same effect as an impact on the whole system.

Fig.~\ref{fig_hist_weak}. shows an identical arrangement except that now \textit{the agent singled out is outside the strongly correlated cluster}. The effect is nil, a field of any strength leaves the rest of the system unperturbed.

The above results are completely natural, they could not have turned out otherwise. If we phrase the message in the language of financial systemic risk, we can say the following. An agent becomes systemically important if it belongs to the strongly correlated cluster. If we wish to identify the systemically important financial institutions, then, in addition to the largest and most connected banks, we have to identify the strongly \textit{correlated} banks. Note that connectedness and correlations are very different notions: an agent can be connected with every other one, yet essentially uncorrelated with them, if the influences of her peers happen to cancel out on her. Our model is constructed such that no agent can be ``too large to fail'', because all of them are equivalent. No agent can be ``too interconnected to fail'' either, or more precisely all of them are too interconnected, because, sitting on the complete graph, all of them are interconnected with every other one. The only difference between them is in their position in the network of interactions. If they become strongly correlated with a large number of their partners, their good or bad luck will be shared by most of the others. If, however, they decouple from the rest, hardly anyone will notice whatever happens to them.

At a recent Latsis Symposium Joseph \citet{24} mentioned the property ``too correlated to fail'' as one of the important criteria for systemically important financial institutions. Our results above are an illustration of this idea. What we can add to this notion on the basis of our present studies is that the property of ``too much correlatedness'' is not an intrinsic property of the \textit{agent}, it is an intrinsic property of the \textit{system}, because a slight change in the interaction matrix may remove the agent from the cluster of strongly correlated partners, but will not annihilate the cluster itself. 

Our work shows that strongly correlated clusters emerge for any realization of the interaction matrix (except in the cases when all, or almost all the couplings are negative). In the interaction-dominated regime this cluster will always contain a large part of the system. The precise conditions for an agent to belong to this cluster or falling outside depend not only on its interactions with the others, but also on the noise and external field. Just as the phase space landscape reacts very sensitively to slight changes in the control parameters, so will the correlations. A strongly correlated large cluster may appear to be extremely stable and solid. Yet it poses a high level of systemic risk, precisely because the correlations can hugely amplify any small shift in the status of even a small number of its constituents.

\section{The distribution of correlations on a two-dimensional regular lattice} 
\label{sec_lattice}

Now we turn to the behaviour of correlations in the same heterogeneous agent model, this time with the agents situated at the nodes of a regular two-dimensional ($2d$) square lattice.

One may wonder whether the tendency for strongly correlated clusters to emerge on the complete graph is directly linked to the extreme connectedness of that network. We shall see shortly that this is not the case: strongly correlated clusters can easily build up also on a $2d$ lattice, provided that the noise level is low enough. 

We have measured the histogram of correlations also in the 2d system, and the results are qualitatively not much different from those presented in Figs. 8 to 12. Instead of repeating them here, we have chosen a different representation of the basic phenomenon of extended correlations. In finite dimensional regular lattices we have an evident measure of distance, and we can study not only the histogram of the correlations, but their spatial extent as well.

\begin{figure}[H] 
  \centering
  \includegraphics[width=\columnwidth]{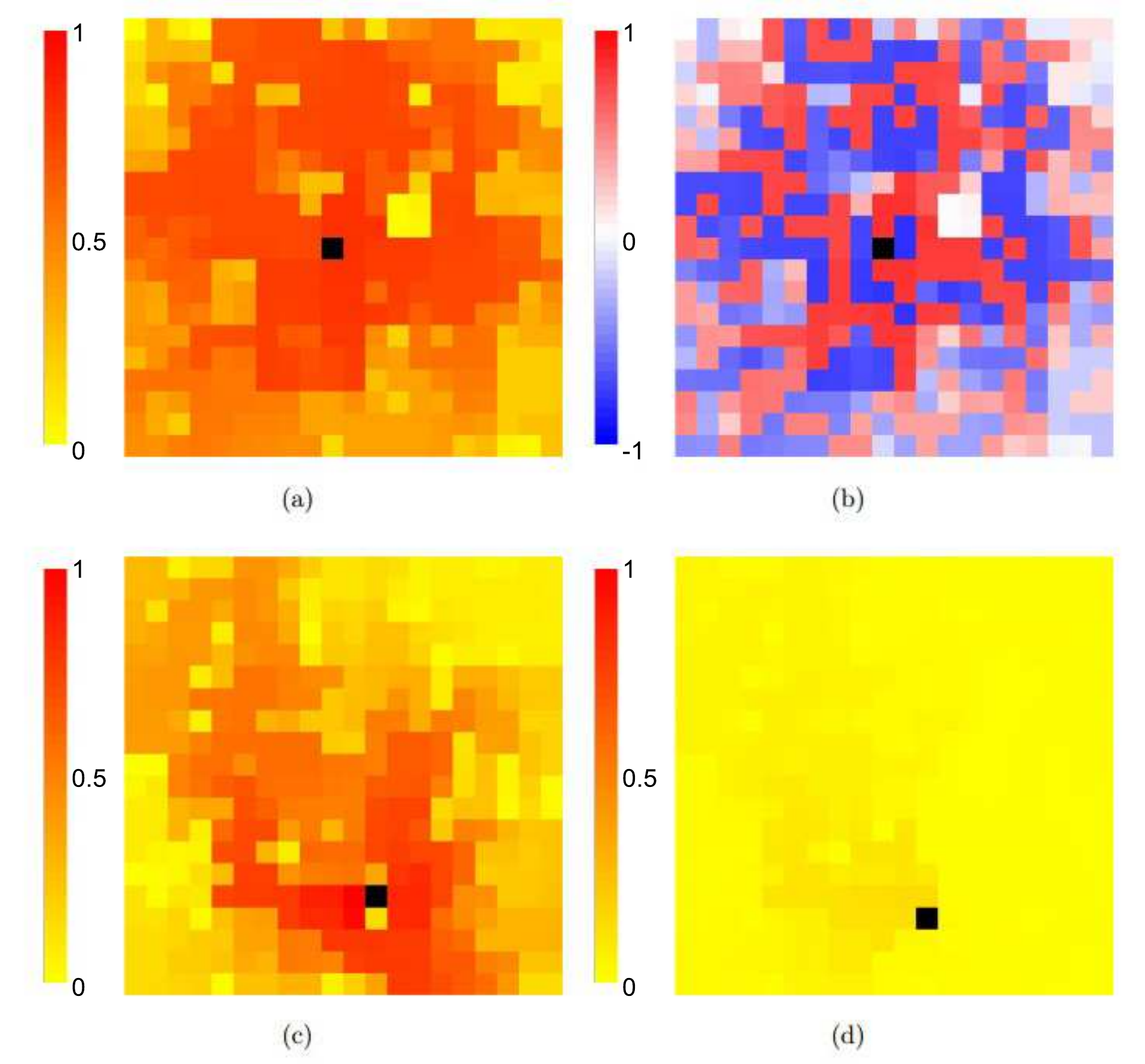}
  \caption{Distribution of correlations over a lattice of $20\times20$, measured from the agents indicated in black. Free boundary conditions were applied. ($a$) The distribution of absolute values. Colour code: deep red: values close to $1$, light yellow: values close to $0$. ($b$) Distribution of the same correlations with sign. Colour code: deep blue: values close to $-1$, deep red: values close to $+1$, light colours: values around zero. ($c$) Absolute values of correlations measured from a different agent. ($d$) Same as in ($c$), measured from a next neighbour of ($c$), the one right below.}
 \label{fig_corr_2d}
\end{figure}

Fig.~\ref{fig_corr_2d}($a$) shows a typical example of the distribution over the lattice of the absolute values of correlations, measured from the agent indicated in black at the middle, in a $20\times20$ sample. This particular agent is a member of a strongly correlated cluster that spans the whole sample. The colours code the absolute values, ranging from zero (yellow) to one (deep red).

Fig.~\ref{fig_corr_2d}($b$) is the same with the signs also coded, red meaning positive, blue negative. In addition to the extent of this strongly correlated cluster, a remarkable feature of Fig.~\ref{fig_corr_2d}($b$) is that positive correlations exist even between partners that cannot be linked by purely red paths. The logic of ''a friend's friend is a friend'' and ''an enemy's enemy is a friend'' are both at work here.

Fig.~\ref{fig_corr_2d}($c$) shows the distribution of the absolute values measured from a different reference agent. As can be seen, this agent has fewer friends than the one indicated in Fig.~\ref{fig_corr_2d}($a$). The agent marked black in Fig.~\ref{fig_corr_2d}($d$) is the nearest neighbour of the previous one. This is a lonely agent: it is outside the strongly correlated cluster. This example shows that an agent can be very weakly correlated with the others, even if it is directly linked to a strongly correlated agent.

These simulations were performed at noise level $T=0.6$, by averaging over Monte Carlo runs of length $10^6$ sweeps. As will be seen in the next section, the boundary conditions play an important role in these strongly correlated systems; in Fig.~\ref{fig_corr_2d}($a$) to ($d$) free boundary conditions were applied.

We note that we also performed extended simulations of three-dimensional samples with very similar results: large, strongly correlated clusters always emerge when the noise level is not two high. They span the whole system, with the rest of the agents scattered over the lattice at random.  

\section{The role of boundary conditions}
\label{sec_boundary}

Simulations of assemblies of interacting entities in finite dimensions have to specify the boundary conditions. If the interactions are homogeneous (translationally invariant) it is well justified to apply periodic boundary conditions, in order to get rid of the effect of agents occupying special positions. Periodic boundary conditions tend to smooth out finite size fluctuations and facilitate the extrapolation to very large system sizes. As such they enjoy wide spread popularity, and are deeply ingrained into the culture of simulations.

In view of the long range random correlations we have just described, simulations of heterogeneous, translationally non-invariant systems should exercise a lot more caution in choosing the right boundary conditions, however. The large, strongly correlated spanning clusters transmit any impact on the boundary right across the whole system. When one is imposing periodic boundary conditions, one is inadvertently reorganizing also the whole system. In this section we are going to demonstrate how different boundary conditions may reorganize the pattern of correlations deep inside the system, and how any averaging (either over the lattice, or over the samples) can mask this effect.

\begin{figure}[H] 
  \centering
  \includegraphics[width=\columnwidth]{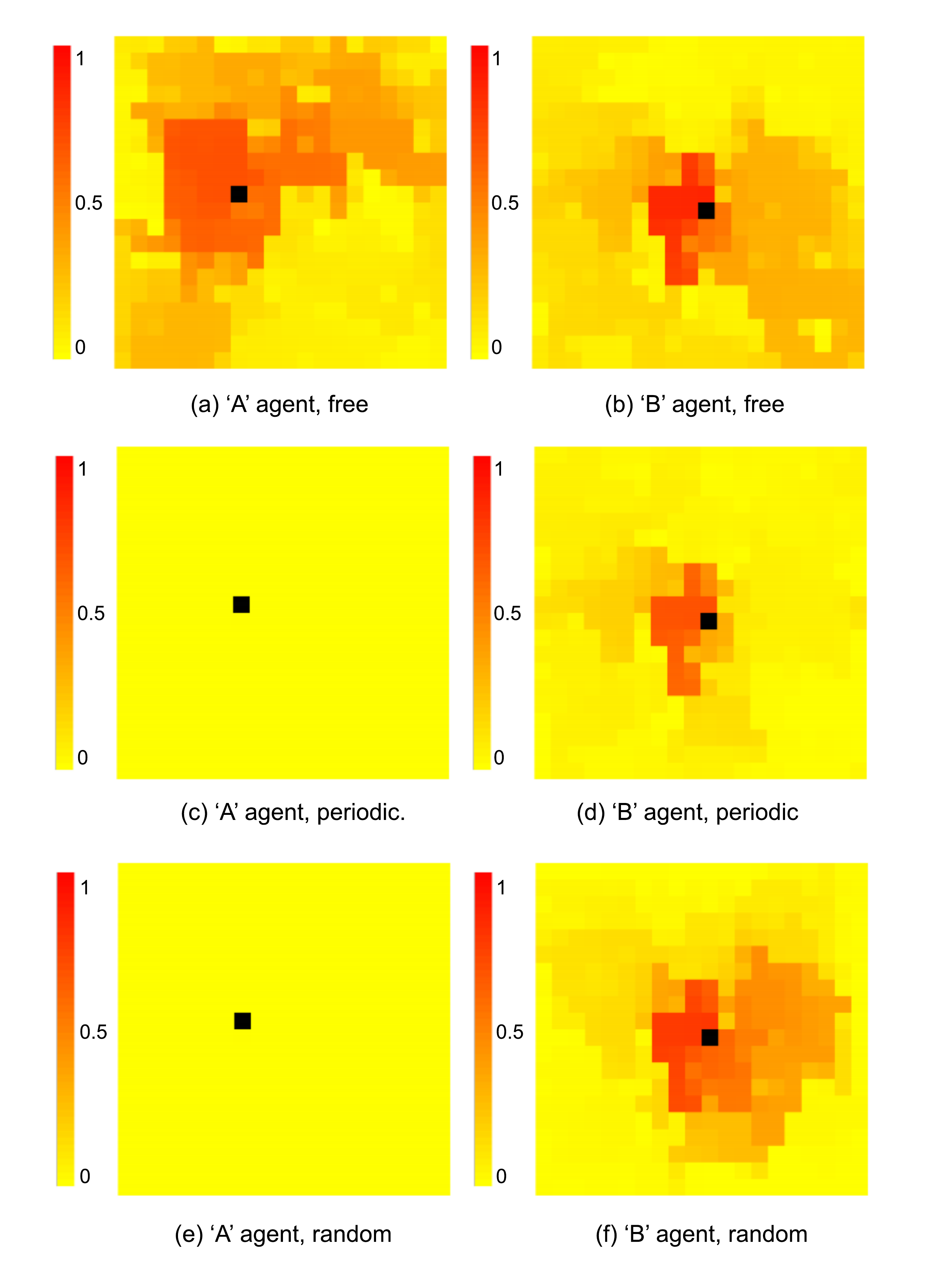}
  \caption{The effect of different boundary conditions on the spatial distribution of correlations relative to two different agents in a $20\times20$ system. The noise level is $T=0.6$, the results are averaged over $10^6$ MC sweeps. The colour code is the same as before. See main text for more explanation.}
 \label{fig_corr_bound}
\end{figure}

Fig.~\ref{fig_corr_bound} gives an indication of the effect different boundary conditions can have on the structure of the system. We are presenting a given $20\times20$ sample here, with the spatial distribution of correlations relative to two different reference agents shown in the left column, ($a$), ($c$), ($e$), and the right column, ($b$), ($d$), ($f$), respectively. The first line, ($a$) and ($b$) was obtained with free boundary conditions, the second, ($c$) and ($d$), with periodic boundary conditions, and the third, ($e$) and ($f$), with random boundary conditions. (Random boundary conditions mean embedding the system into one more layer of agents fixed at randomly chosen $\pm1$ values.)

As can clearly be seen, the effect of changing the boundary conditions can be radical. The strongly correlated cluster around the reference agent in Fig.~\ref{fig_corr_bound}($a$) is annihilated by imposing either periodic ($c$), or random ($e$) boundary conditions. In contrast, the cluster in Fig.~\ref{fig_corr_bound}($b$) gets only reduced by periodic boundary conditions ($d$), and rearranged by random ones ($f$). It is clear that the effect of boundary conditions on one or another property of a heterogeneous system will be essentially unforeseeable; this effect depends on so many details that basically any rearrangement can take place.

We have performed extensive studies on a large number of similar samples in two dimensions, and also on three dimensional samples of size $10^3$, at $T=1.0$, averaged over $2\times10^7$ MC sweeps, with the same general result as those above. In addition, we also followed the behaviour of correlations along specific lines, obtaining essentially random functions with a weak tendency to decay, but randomly jumping up and down along the way, with a rather large probability of taking on $\mathcal{O}(1)$ values even at the largest distances permitted by our small lattices. We do not believe that displaying all the corresponding figures would add much to what we have already presented; the overall picture must be fairly clear by now.

We would like to emphasize that all the above observations were made on individual samples, that is on systems with a given realization of the interaction matrix. As discussed in the Introduction, in spin glass research it is customary to average over the whole set of realizations of the random couplings. We believe that, because of the mesoscopic scale of multiagent systems, averaging over random couplings cannot be justified. Nevertheless, let us investigate what would happen to the boundary condition dependence on average. We shall see that averaging the wildly fluctuating correlations over a relatively small number of interaction samples makes them insensitive to boundary conditions.
 
\begin{figure}[H] 
  \centering
  \includegraphics[width=\columnwidth]{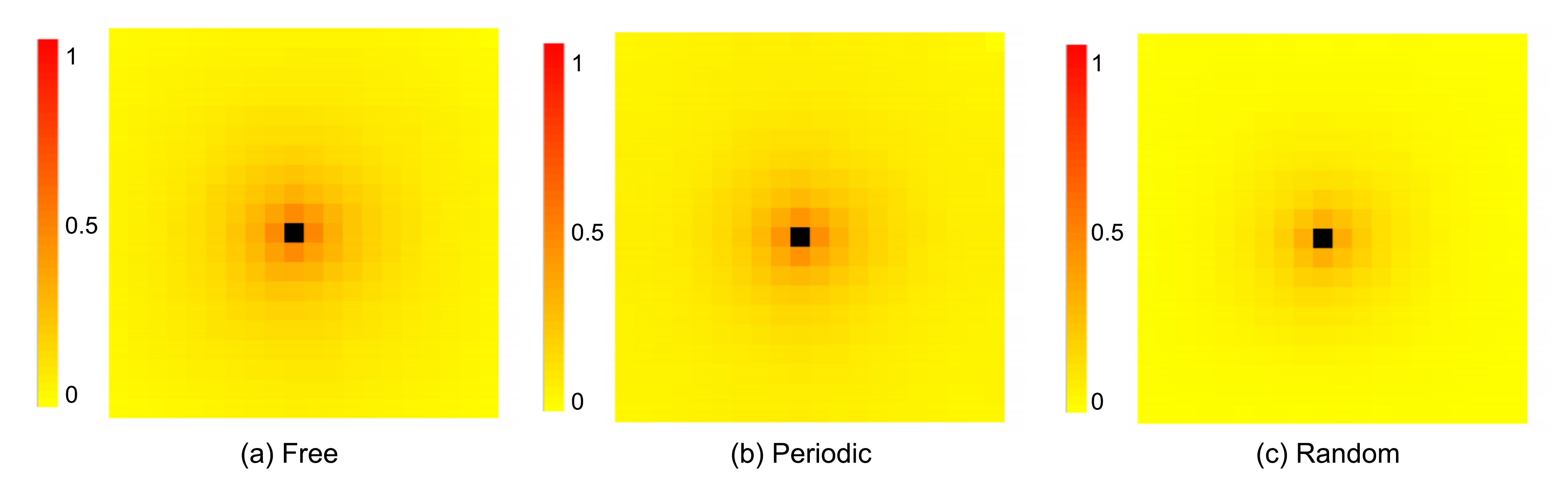}
  \caption{Spatial distribution of correlations on the same type of lattice as in Fig.~\ref{fig_corr_bound}, after averaging over $500$ randomly chosen realizations of the interaction matrix. The results suggest a smooth spatial dependence and independence of the boundary conditions.}
 \label{fig_corr_bound_avg}
\end{figure}

Let us look at Fig.~\ref{fig_corr_bound_avg}. It shows the spatial distribution of correlations on the same type of lattice we have already seen in Fig.~\ref{fig_corr_bound} (size $20\times20$, $T=0.6$, correlations averaged over $10^6$ sweeps), with the only difference that here the results have been averaged over $500$ randomly chosen samples of the model. Sample averaging evidently smooths out the results and make them largely independent of the boundary conditions. The sample averaged correlations fall off smoothly and monotonically. The small size of our lattice does not allow us to determine the character (exponential or power law) of the averaged correlations.

The results obtained in $3d$ are very similar, and will not be presented here.

The observation that sample averaging smooths out the boundary condition dependence leaves a nagging question behind. Real world systems exist in one given sample. We would like to be able to say something reliable about the correlations in that concrete sample, something that does not depend so drastically on the boundary conditions, without invoking averaging over the possible alternative realizations of the same system. The following consideration may provide a clue.

As we know from the theory of spin glasses, \citet{5}, macroscopic aggregates (e.g. the sum of agents' votes or the time average of cost)  self-average, become independent of the concrete sample, therefore they obviously cannot display this sort of irregular behaviour. In particular, they must have a proper large size limit, and cannot be so sensitive to boundary conditions. The local correlations we have considered here do depend on the boundary conditions and, at least within our small systems sizes, do not settle down to a limiting value. It is plausible to expect, however, that upon some kind of coarse graining (partial aggregation) the wild fluctuations get attenuated: there exist an intermediate scale where the dependence on boundary conditions and other small details becomes weaker, perhaps even negligible.  An interesting open question is then to what degree of aggregation should we go up to, in order to get rid of the nuisance of the drastic dependence on the boundary conditions.

What we can offer is a partial answer. Averaging over parallel histories  is not the only way to tame the strong fluctuations and boundary condition dependence. To achieve the same goal, it is sufficient to average the correlations over a ring around the reference agent. As our system is small, the rings drawn about the reference agent do not contain too many other agents, yet an averaging over these small sets is sufficient to smooth out the correlation functions as efficiently as an averaging over the random samples.

The effect of averaging over spherical shells in $3d$ is entirely similar to the above.

\section{Summary}
\label{sec_summary}

Let us briefly summarize the most important ideas we have presented and tried to illustrate with the example of a spin-glass-like multiagent model in this paper.

The first fundamental property of these systems is their very slow, punctuated equilibrium type dynamics that makes the observed structure dependent on the time length of observation, but also allows one to make meaningful measurements of observables within the life time of these long lived quasi-equilibria.

The network of couplings does not directly determine the dynamics. The dynamics is governed by a functional, the cost function, defined over the network of couplings. The microscopic states define another network in phase space, and the dynamics is a random walk over this phase space graph. In the low cost regime only the low lying states matter, and the cost landscape is very rugged, displaying several valleys separated by barriers of varying heights. The landscape is related to the couplings in a highly nontrivial way, sometimes a slight rearrangement of the network of couplings induces a strong rearrangement of the attractors and their basins. There are typically many attractors, so the dynamics is chaotic, the trajectories strongly depend on initial conditions. Moreover, the structure of this attractor landscape also depends chaotically on the details of the interaction network and on the control parameters, like the background noise (''temperature'') and the external field. This is not unlike the turbulent evolution of a biosystem in a fitness landscape shaped both by the competitors and also by external factors, or that of an economy with its competing and collaborating actors.

Due to the interactions between the agents, large, strongly correlated clusters emerge in the interaction dominated (low noise) regime for any concrete realization of the interaction matrix. These correlated clusters span the whole system and their structure depends sensitively on all kinds of details again: on the precise distribution of couplings, on control parameters and boundary conditions. They make the system \textit{nonlocal}, an impact on an agent belonging to the strongly correlated cluster is transmitted to the whole system. Clearly, these strongly correlated agents are the systemically important ones. Unfortunately, their identification is extremely difficult, because the structure of the strongly correlated clusters depends on the full distribution of the couplings over the system. The implications of this state of affairs for the problem of financial systemic risk are obvious.

We have also studied the spatial distribution of correlations in two\--di\-men\-si\-o\-nal regular lattice. We have found that this distribution shows a very pronounced dependence on the boundary conditions.

All the above statements seem to converge towards a very negative conclusion: these models are extremely unstable with respect to any tiny change in any conceivable factor or detail that has an influence on them, which also implies that there is no hope whatsoever to discover their structure by observations of their behaviour. Nevertheless, the results about averaging presented in Section~\ref{sec_boundary} show some promise of a resolution. We saw there that the dependence on boundary conditions can be cured by averaging over the random samples (which is not much of a consolation, as the real world systems we have in mind exist in one given sample), but we also learned that the same stable results can be achieved by averaging over small regions in the lattice. This means that coarse-graining can take care of the drastic fluctuations. We raised the question to which level of coarse graining one has to go in order to get rid of the instability, or, conversely, to what level of resolution can one disaggregate data, before running into the insurmountable difficulties with the instabilities of high dimensional models. It is obvious that the answer is of fundamental importance for the acceptance of agent based models. If we have to course-grain up to the level of the usual macroeconomic variables then ABM's will dissolve into the mainstream models. If they stop depending on tiny details already at an intermediate level, they will contribute real value to the understanding of economics. A detailed examination of these questions is left for future work.

\section*{Acknowledgement}
We are obliged to Alan Kirman for critically reading the manuscript. This work has been supported by the European Union under grant agreement No. FP7-ICT-255987-FOC-II Project; by the Institute for New Economic Thinking under grant agreement ID: INO1200019; by the European Union and the European Social Fund under grant agreement No. T\'AMOP 4.2.1./B-09/KMR-2010-0003; and by the National Innovation Office under grant No. KCKHA005.


\bibliographystyle{abbrvnat}

\bibliography{work}   
%
%
%


\end{multicols}

\end{document}